\documentclass[12pt,fleqn]{article} \textheight=9in
\newcommand{\ra}{\rightarrow}
\newcommand{\bra}{\langle} \newcommand{\ket}{\rangle}
\newcommand{\be}{\begin{equation}}
\newcommand{\ee}{\end{equation}}
\newcommand{\bea}{\begin{eqnarray}}
\newcommand{\eea}{\end{eqnarray}}
\newcommand{\eps}{\epsilon}
\newcommand{\E}{\mbox{e}}
\newcommand{\e}{\mbox{\scriptsize e}}
\newcommand{\ffi}{\varphi}

\newcommand{\ep}{\qquad {\vrule height 10pt width 8pt depth 0pt}}

\newcommand{\grintl}{[\kern-.18em [}
\newcommand{\grintr}{]\kern-.18em ]}
\newcommand{\ds}{\displaystyle}

\newtheorem{theorem}{Theorem}

\newtheorem{lemma}[theorem]{Lemma}
\newtheorem{corollary}[theorem]{Corollary}

\def\R{\hbox{$\mit I$\kern-.277em$\mit R$}}
\def\C{\hbox{$\mit I$\kern-.6em$\mit C$}}
\def\un{\hbox{$\mit I$\kern-.77em$\mit I$}}
\def\0{\hbox{$\mit I$\kern-.70em$\mit O$}}
\def\r{I\kern-.277em R}

\def\N{\mbox{\bf N}}

\begin{document}
\title{Exponentially Accurate Semiclassical Dynamics: Propagation,
  Localization, Ehrenfest Times, Scattering and More General States}

\author{George A. Hagedorn\thanks{Partially
Supported by National Science Foundation
Grant DMS--9703751.}\\
Department of Mathematics and\\
Center for Statistical Mechanics and Mathematical Physics\\
Virginia Polytechnic Institute and State University\\
Blacksburg, Virginia 24061-0123, U.S.A.\\[15pt]
\and
Alain Joye\\
Institut Fourier\\ Unit\'e Mixte de Recherche CNRS-UJF 5582\\
Universit\'e de Grenoble I\\
BP 74\\
F--38402 Saint Martin d'H\`eres Cedex, France}
\maketitle

\begin{abstract}
We prove six theorems concerning exponentially accurate semiclassical quantum
mechanics. Two of these theorems are known results, but have new proofs.
Under appropriate hypotheses, they conclude
that the exact and approximate dynamics of an initially localized wave 
packet agree up to exponentially small errors in $\hbar$ for finite
times and for Ehrenfest times. Two other theorems
state that for such times the wave packets are localized near a
classical orbit up to exponentially small errors. 
The fifth theorem deals with infinite times and states an
exponentially accurate scattering result. The sixth theorem provides extensions
of the other five by allowing more general initial conditions.
\end{abstract}

\vfill
\newpage

\section{Introduction}
\setcounter{equation}{0}
\setcounter{theorem}{0}

This paper is devoted to proving several theorems concerning exponentially
accurate approximations to solutions of the time-dependent Schr\"odinger
equation
\be\label{pde}
i\,\hbar\,\frac{\partial }{\partial t}\,\Psi(x,t,\hbar)\ =\
-\,\frac{\hbar^2}{2}\,\Delta\,\Psi(x,t,\hbar)\,+\,V(x)\,\Psi(x,t,\hbar)
\ee
in the semiclassical limit $\hbar\ra 0$.

The semiclassical approximation of quantum dynamics has been the
object of several recent investigations from different points of
view. One approach uses coherent state initial conditions
and approximates the evolved wave packet
by suitable linear combinations of coherent states.
Another approach considers the Heisenberg evolution
of suitable bounded observables and approximates the corresponding
operators by means of Egoroff's theorem.
The goal of both approaches is to produce accurate, computable
approximations as $\hbar$ goes to zero, for as long
a time interval as possible. In scattering situations the time interval is
the whole real line.

There are several results concerning the propagation of certain coherent states
for finite time intervals. Early results \cite{hepp,scat} constructed
approximate solutions that were accurate up to $O(\hbar^{1/2})$ errors.
Later approximations were constructed with $O(\hbar^{l/2})$ errors for any $l$
\cite{semi3,semi4,CR,Paul-Uribe}. Very recently, approximations were 
constructed with errors of exponential order $O(\E^{-\Gamma/\hbar})$
with $\Gamma>0$ in \cite{hagjoy3} (see also \cite{Yajima}).

The validity of the corresponding approximations for time intervals
of length $O(\ln(1/\hbar))$, the so-called Ehrenfest time-scale,
has been established up to $O(\hbar^{l/2})$ errors in \cite{CR}, and
up to $O(e^{1/\hbar^{\alpha}})$ errors with $0<\alpha<1$ in \cite{hagjoy3} .
There is physical intuition and evidence that the Ehrenfest time scale is the
natural limit for the validity of coherent state type approximations.
This issue is studied in detail for the quantized Baker and Cat maps
in \cite{bdb}.

Approximations have been constructed for infinite times in the context of
scattering theory for coherent states. Approximate solutions with errors of
order $O(\hbar^{1/2})$, uniformly in time, are produced in \cite{scat}. This
yields approximations for the scattering matrix with errors that are also
$O(\hbar^{1/2})$. Related results for another class of states can be found in
\cite{Y1,Y2}.

Corresponding results for the approximation
of observables in the Heisenberg picture can be found in
\cite{Robert} for approximations with $O(\hbar^{l/2})$  errors for any $l$ for
finite times. Approximations with exponentially small errors both for finite
times and for Ehrenfest times are constructed in \cite{BGP} and \cite{br}.

The exponentially accurate results mentioned above, and those we present below,
are obtained for Hamiltonians that satisfy certain analyticity conditions.
The approximations are generated by optimal truncation of asymptotic series.

For further information, we refer the reader to the review articles
\cite{c,robert,Paul,gituab}.

\vskip .5cm
The present paper is concerned with the propagation of coherent
states in the spirit of the first approach described above.
We present a new construction of approximate solutions
to the time dependent Schr\"odinger equation that is an alternative to the one
presented in \cite{hagjoy3}.

The new expansion has several advantages. In addition to being
exponentially accurate up to the Ehrenfest time scale (Theorems
\ref{thm1} and \ref{thm3}), it allows
us to extend our previous results in four separate directions:\\
1.\quad We get exponentially precise localization properties for
both the approximation and the exact solution for both finite times and
Ehrenfest times (Theorems \ref{thm2} and \ref{thm4}).\\
2.\quad We get exponentially accurate information on the
semiclassical limit of the scattering matrix for suitable short range
potentials (Theorem \ref{thm5}).\\
3.\quad The new algorithm is superior for numerical computation.
The work done to construct the approximate wave function for one value
of $\hbar$ is used for the construction for all smaller values of $\hbar$.
This should be contrasted with the construction of \cite{hagjoy3}
where every calculation must be redone for each value of $\hbar$.\\
4.\quad The results in \cite{hagjoy3} concern the propagation of
initial coherent states given by a linear combination of a finite
number $N$ of elementary coherent states. We can control
the new approximation as a function of $N$, which
also allows us to extend the validity of all previous results to
a more general set of initial states (Theorem \ref{thm6}).
In this case however, the algorithm requires the computation of different
quantities as $\hbar$ varies.

\vskip .25cm
The technical difference between the present construction and
the one in \cite{hagjoy3} is the following:\,\
In both papers, we use a suitable time dependent basis to 
convert the PDE (\ref{pde}) into an infinite system of ODE's for the
expansion coefficients in that basis of the solution to the Schr\"odinger 
equation. In \cite{hagjoy3} we then construct the approximate solution by 
approximating this infinite system by a finite system, which we solve
exactly.
In the new approach, we substitute an {\it a priori}
expansion in powers of $\hbar^{1/2}$ into the original infinite system
of ODE's.
We construct our approximate solution by keeping a finite number of terms.
This turns out to be quite efficient.

\vskip .25cm
The new approximation also plays a vital role in the construction of
an exponentially accurate time--dependent Born--Oppenheimer approximation
\cite{hagjoy5}.

\vskip .5cm
\section{Coherent States and Classical Dynamics}\label{wp}
\setcounter{equation}{0}

\vskip .25cm
We begin this section by recalling the definition of the coherent states
$\phi_j(A,\,B,\,\hbar,\,a,\,\eta,\,x)$ described in detail in \cite{raise}.
A more explicit, but more complicated definition is given in \cite{semi4}.

We adopt the standard multi-index notation. A multi-index
$j=(j_1,\,j_2,\,\dots ,\,j_d)$ is a $d$-tuple of non-negative integers.
We define
%
%
$|j|=\sum_{k=1}^d\,j_k$,\ \,
$x^j=x_1^{j_1}x_2^{j_2}\cdots x_d^{j_d}$,\\
$j!=(j_1!)(j_2!)\cdots(j_d!)$,\ \,
and\ \,
$D^j=\frac{\partial^{|j|}}{(\partial x_1)^{j_1}(\partial x_2)^{j_2}\cdots
(\partial x_d)^{j_d}}$.

Throughout the paper we assume $a\in\R^d$, $\eta\in\R^d$ and $\hbar>0$. We
also assume that $A$ and $B$ are $d\times d$ complex invertible matrices that
satisfy
\bea\label{cond1}
A^t\,B\,-\,B^t\,A&=&0,\nonumber \\
A^*\,B\,+\,B^*\,A&=&2\,I.
\eea

These conditions guarantee that both the real and imaginary parts of $BA^{-1}$
are symmetric. Furthermore, $\mbox{Re}\,BA^{-1}$ is strictly positive definite,
and $\left(\mbox{Re}\,BA^{-1}\right)^{-1}=\,A\,A^*$.

Our definition of $\ffi_j(A,\,B,\,\hbar,\,a,\,\eta,\,x)$ is based on the
following raising operators that are defined for $m=1,\,2,\,\dots ,\,d$.
$$
\mathcal{ A}_m(A,B,\hbar,a,\eta)^*
\ =\
\frac{1}{\sqrt{2\hbar}}\,\left[\,\sum_{n=1}^d\,\overline{B}_{n\,m}\,(x_n-a_n)
\ -\,i\ \sum_{n=1}^d\,\overline{A}_{n\,m}\,
(-i\hbar\frac{\partial\phantom{x^n}}{\partial x_n}-\eta_n)\,\right] .
$$
The corresponding lowering operators $\mathcal{ A}_m(A,B,\hbar,a,\eta)$ are
their formal adjoints.

These operators satisfy commutation relations that lead to the properties of
the\\
$\phi_j(A,\,B,\,\hbar,\,a,\,\eta,\,x)$ that we list below.
The raising operators $\mathcal{ A}_m(A,B,\hbar,a,\eta)^*$ for\\
$m=1,\,2,\,\dots,\,d$ commute with one another, and the lowering operators
$\mathcal{A}_m(A,B,\hbar,a,\eta)$ commute with one another. However,
$$\mathcal{A}_m(A,B,\hbar,a,\eta)^*\,\mathcal{A}_n(A,B,\hbar,a,\eta)\,-\,
\mathcal{A}_n(A,B,\hbar,a,\eta)\,\mathcal{A}_m(A,B,\hbar,a,\eta)^*
\ =\ \delta_{m,\,n}.$$

\vskip .25cm
\noindent
{\bf Definition}\quad
For the multi-index $j=0$, we define the normalized complex Gaussian wave packet
(modulo the sign of a square root) by
\bea\nonumber
&&\phi_0(A,\,B,\,\hbar,\,a,\,\eta,\,x)\,=\,\pi^{-d/4}\,
\hbar^{-d/4}\,(\det(A))^{-1/2}\\[6pt]\nonumber
&&\qquad\qquad\quad \times\quad
\exp\left\{\,-\,\langle\,(x-a),\,B\,A^{-1}\,(x-a)\,\rangle/(2\hbar)\,
+\,i\,\langle\,\eta,\,(x-a)\,\rangle/\hbar\,\right\} .
\eea
Then, for any non-zero multi-index $j$, we define
\bea\nonumber
\phi_j(A,\,B,\,\hbar,\,a,\,\eta,\,\cdot\,)&=&
\frac{1}{\sqrt{j!}}\
\left(\,\mathcal{A}_1(A,B,\hbar,a,\eta)^*\right)^{j_1}\
\left(\,\mathcal{A}_2(A,B,\hbar,a,\eta)^*\right)^{j_2}\ \cdots \\[5pt]
\nonumber &&\qquad\qquad\qquad\ \times\
\left(\,\mathcal{A}_d(A,B,\hbar,a,\eta)^*\right)^{j_d}\
\phi_0(A\,,B,\,\hbar,\,a,\,\eta,\,\cdot\,).
\eea

\vskip .25cm
\noindent
{\bf Properties}\quad
1.\quad For $A=B=I$, $\hbar=1$, and $a=\eta=0$, the
$\phi_j(A,\,B,\,\hbar,\,a,\,\eta,\,\cdot\,)$ are just the standard Harmonic
oscillator eigenstates with energies $|j|+d/2$.\\[7pt]
2.\quad For each admissible $A$, $B$, $\hbar$, $a$, and $\eta$,
the set $\{\,\phi_j(A,\,B,\,\hbar,\,a,\,\eta,\,\cdot\,)\,\}$
is an orthonormal basis for $L^2(\R^d)$.\\[7pt]
3.\quad The raising operators can also be given by another formula that was
omitted from \cite{raise} in the multi-dimensional case.
If we set
$$g(A,\,B,\,\hbar,\,a,\,x)\,=\,
\exp\left\{\,-\,\langle\,(x-a),\,
\left(BA^{-1}\right)^*(x-a)\,\rangle/(2\hbar)\,-\,i
\langle\,\eta,\,(x-a)\,\rangle/\hbar\,\right\} ,$$
then we have
\bea\nonumber
&&\left(\,\mathcal{ A}_m(A,B,\hbar,a,\eta)^*\,\psi\,\right)(x)\\[4pt]
&&\qquad\qquad =\ -\ \sqrt{\frac{\hbar}2}\ \frac 1{g(A,\,B,\,\hbar,\,a,\,x)}\
\sum_{n=1}^d\ \overline{A}_{n\,m}\,
\frac{\partial\phantom{x^n}}{\partial x_n}\,
\left(\,g(A,\,B,\,\hbar,\,a,\,x)\,\psi(x)\,\right) .\nonumber
\eea
4.\quad In \cite{semi4}, the state $\phi_j(A,\,B,\,\hbar,\,a,\,\eta,\,x)$ is
defined as a normalization factor times
$$
\mathcal{ H}_j(A;\,\hbar^{-1/2}\,|A|^{-1}\,(x-a))\
\phi_0(A,\,B,\,\hbar,\,a,\,\eta,\,x).
$$
Here $\mathcal{ H}_j(A;\,y)$ is a recursively defined
$|j|^{\mbox{\scriptsize th}}$
order polynomial in $y$ that depends on $A$ only through
$U_A$, where $A=|A|\,U_A$ is the polar decomposition of $A$.\\[7pt]
5.\quad By scaling out the $|A|$
and $\hbar$ dependence and using Remark 3 above, one can show that
$\mathcal{ H}_j(A;\,y)\,\E^{-y^2/2}$ is an (unnormalized) eigenstate of the usual
Harmonic oscillator with energy $|j|+d/2$.\\[7pt]
6.\quad When the dimension $d$ is $1$, the position and momentum uncertainties
of the\newline
$\phi_j(A,\,B,\,\hbar,\,a,\,\eta,\,\cdot\,)$ are
$\sqrt{(j+1/2)\hbar}\ |A|$ and $\sqrt{(j+1/2)\hbar}\ |B|$, respectively. In
higher dimensions, they are bounded by
$\sqrt{(|j|+d/2)\hbar}\ \|A\|$ and $\sqrt{(|j|+d/2)\hbar}\ \|B\|$,
respectively.\\[7pt]
7.\quad When we approximately solve the Schr\"odinger equation, the choice of
the sign of the square root in the definition of
$\phi_0(A,\,B,\,\hbar,\,a,\,\eta,\,\cdot\,)$ is determined by continuity in $t$
after an arbitrary initial choice.\\[7pt]
8.\quad We prove below that the matrix elements of $(x-a)^m$ satisfy
\bea\nonumber
&&\left|\,\langle\,\phi_j(A,\,B,\,\hbar,\,a,\,\eta,\,x),\,(x-a)^m\,
\phi_k(A,\,B,\,\hbar,\,a,\,\eta,\,x)\rangle\,\right|\\[7pt]\nonumber
&\leq&\hbar^{|m|/2}\,(\sqrt{2}d)^{|m|}\,\| A\|^{|m|}\,
\sqrt{(|k|+1)(|k|+2)\cdots (|k|+|m|)},
\eea
and
\bea\nonumber
\langle\,\phi_j(A,\,B,\,\hbar,\,a,\,\eta,\,x),\,(x-a)^m\,
\phi_k(A,\,B,\,\hbar,\,a,\,\eta,\,x)\,\rangle\,=\,0,
&\mbox{if}&\bigg| |j|-|k|\bigg| >|m|.
\\ &&\label{esgaus}
\eea

We now assume that the potential $V:\R^d\ra \R$ is smooth and
bounded below. Our semiclassical approximations depend on solutions to the
following classical equations of motion
\bea
\dot{a}(t)&=&\eta(t) ,\nonumber \\
\dot{\eta}(t)&=&-\,\nabla V(a(t)),\nonumber \\
\dot{A}(t)&=&i\,B(t),\label{newton} \\
\dot{B}(t)&=&i\,V^{(2)}(a(t))\,A(t),\nonumber \\
\dot{S}(t)&=&\frac{\eta(t)^2}2\,-\,V(a(t)),\nonumber
\eea
where $V^{(2)}$ denotes the Hessian matrix for $V$, and the
initial conditions $A(0)$, $B(0)$, $a(0)$, $\eta(0)$, and
$S(0)=0$ satisfy (\ref{cond1}).

The matrices $A(t)$ and $B(t)$ are related to the linearization of the
classical flow through the following identities:
\bea
A(t)&=&\frac{\partial a(t)}{\partial a(0)}\,A(0)\,+\,
i\,\frac{\partial a(t)}{\partial \eta(0)}\,B(0),\nonumber\\[4pt]\nonumber
B(t)&=&\frac{\partial\eta(t)}{\partial \eta(0)}\,B(0)
\,-\,i\frac{\partial\eta(t)}{\partial a(0)}\,A(0).
\eea

Because $V$ is smooth and bounded below, there exist global solutions to
the first two equations of the system (\ref{newton}) for any initial
condition. From this, it follows immediately that the remaining three
equations of the system (\ref{newton}) have global solutions.
Furthermore, it is not difficult \cite{semi3,semi4} to prove
that conditions (\ref{cond1}) are preserved by the flow.

The usefulness of our wave packets stems from the following important
property \cite{raise}.
If we decompose the potential as
\begin{equation}\label{defy}
V(x)\,=\,W_{a}(x)\,+\,Z_a(x)\,\equiv W_{a}(x)\,+\,(V(x)-W_a(x)),
\end{equation}
where $W_a(x)$ denotes the second order Taylor
approximation (with the obvious abuse of notation)
$$
W_a(x)\,\equiv\,V(a)\,+\,V^{(1)}(a)\,(x-a)\,+\,V^{(2)}(a)\,(x-a)^2/2.
$$
then for all multi-indices $j$,
\bea\label{profi}
&&i\,\hbar\,\frac{\partial }{\partial t}\,
\left[\,\E^{iS(t)/\hbar}\,\phi_j(A(t),B(t),\hbar,a(t),\eta(t),x)\,\right]
\nonumber\\[8pt]
&=&\left(\,-\,\frac{\hbar^2}{2}\,\Delta\,+\,W_{a(t)}(x)\,\right)\,
\left[\,\E^{iS(t)/\hbar}\,\phi_j(A(t),B(t),\hbar,a(t),\eta(t),x)\,\right] ,
\eea
if $A(t)$, $B(t)$, $a(t)$, $\eta(t)$, and $S(t)$ satisfy (\ref{newton}).
In other words, our semiclassical wave packets $\ffi_j$ exactly
take into account the kinetic
energy and quadratic part $W_{a(t)}(x)$ of the potential when
propagated by means of the classical flow and its linearization around the
classical trajectory selected by the initial conditions.

In the rest of the paper, whenever we write
$\phi_j(A(t),B(t),\hbar,a(t),\eta(t),x)$, we tacitly assume that
$A(t),B(t),a(t),\eta(t)$, and $S(t)$ are solutions to
(\ref{newton}) with initial conditions satisfying (\ref{cond1}).

\vskip .5cm
\section{The Main Results}
\setcounter{equation}{0}
\setcounter{theorem}{0}

\vskip .25cm
In this section, we list our results concerning the propagation of semiclassical
wave packets. The first is the construction of an approximate wave function that
agrees with the exact wave function up to an exponentially small error.
The construction is quite explicit. It depends on the somewhat arbitrary choice
of a parameter $g>0$.

\vskip .1cm
The precise result is summarized in the following theorem:

\vskip .5cm
\begin{theorem}\label{thm1}
Suppose $V(x)$ is real and bounded below for
$x\in \R^d$. Assume $V$ extends to an analytic function in a neighborhood of
the region
$S_\delta\,=\,\{\,z\,:\,|\,\mbox{Im }z_j\,|\le \delta\,\}$ and satisfies
$|V(z)|\,\le\,M\,\exp\left(\tau|z|^2\right)$ for $z\in S_\delta$ and
some positive constants $M$ and $\tau$.

Fix $T$, choose a classical orbit $a(t)$ for $0\le t\le T$, and
consider an arbitrary normalized coherent state of the form
$$
\psi(x,0,\hbar)=\sum_{|j|\le J}\,c_j(0)\,
\phi_j(A(0),B(0),\hbar,a(0),\eta(0),x).
$$
There exists a number $G>0$, such that
for each choice of the parameter $g\in(0,\,G)$,
there exists an exact solution to the Schr\"odinger equation,
$$i\,\hbar\,\frac{\partial \Psi}{\partial t}\ =\ -\,\frac{\hbar^2}{2}
\,\Delta\,\Psi\ +\, V\,\Psi,
$$
with $\Psi(x,0,\hbar)=\psi(x,0,\hbar)$,
that agrees with the approximate solution
$$\psi(x,t,\hbar)\ =\ e^{iS(t)/\hbar}\,
\sum_{|j|\le J+3g/\hbar-3}\,
c_j(t,\hbar)\,\phi_j(A(t),B(t),\hbar,a(t),\eta(t),x),
$$
up to an error whose $L^2(\R^d)$ norm is bounded by
$\ds C\exp\,\left\{\,-\gamma_g/\hbar\,\right\}$, with $\gamma_g>0$.
Furthermore, the complex coefficients $c_j(t,\hbar)$ are determined by
an explicit procedure.
\end{theorem}

The second result shows that the approximate wave function of Theorem \ref{thm1}
is concentrated within an arbitrarily small distance of the classical
path up to an exponentially small
error if $g$ is chosen sufficently small.

\vskip .5cm
\begin{theorem}\label{thm2}
Suppose that the hypotheses of Theorem \ref{thm1} are satisfied
and that $b>0$ is given. For sufficiently small values of the parameter
$g>0$, the wave packet $\psi(x,t,\hbar)$ is localized within a distance $b$
of $a(t)$, up to an error
$\ds\exp\,\left\{\,-\Gamma_g/\hbar\,\right\}$,
with $\Gamma_g>0$, in the sense that
$$
\left(\,\int_{|x-a(t)|>b}\ |\psi(x,t,\hbar)|^2\,dx\,\right)^{1/2}
\ \le\ \exp\,\left\{\,-\Gamma_g/\hbar\,\right\}.
$$
\end{theorem}

\vskip .3cm
Next, we turn to the validity of the approximation and its localization
properties on the Ehrenfest time scale, i.e. when $T$ is allowed to
increase with $\hbar$ as $\ln(1/\hbar)$.

\vskip .5cm
\begin{theorem}\label{thm3}
Suppose the assumptions of Theorem \ref{thm1} are satisfied except that the
upper bound on $V$ is replaced by  $|V(z)|\,\le\,M\,\exp\left(\tau|z|\right)$
for $z\in S_\delta$ and some positive constants $M$ and $\tau$.
Further, assume the existence of a constant $N>0$ and a positive
Lyapunov exponent $\lambda$ so that $\|A(t)\|\, \leq \, N \exp(\lambda t)$,
for all $t\geq 0$. Then, for sufficiently small $T'>0$, there exist
constants $C'>0$,$\gamma'>0$, $\sigma\in(0,\,1)$, and $\sigma'\in(0,\,1)$,
and an exact
solution to the Schr\"odinger equation that agrees with the approximation
$$\psi(x,t,\hbar)\ =\ e^{iS(t)/\hbar}\,
\sum_{|j|\le J+3/\hbar^{\sigma'}-3}\,c_j(t,\hbar)\,
\phi_j(A(t),B(t),\hbar,a(t),\eta(t),x),
$$
up to an error whose norm is bounded by
$ C'\exp\,\left\{\,-\gamma'/\hbar^{\sigma}\,\right\}$, whenever
$0\leq t\leq T'\ln(1/\hbar)$.\\
Moreover, if $\tau$ can be taken arbitrarily small, we can chose
$T'=\frac{1}{6 \lambda}(1-\eps)$ where $\eps$ is arbitrarily small.
\end{theorem}

\noindent
{\bf Remark:}\quad The semiclassical approximation of observables
in the Heisenberg picture holds for any $T'<2/(3\lambda)$, when
$\tau <<1$, as shown recently in \cite{br}. That time interval is
longer than those for which a localized coherent state can approximate
the evolution of an initial coherent state, which is characterized by
$T'<1/(2\lambda)$. See \cite{bdb} for a study of related issues
on quantized hyperbolic maps on the torus.

\begin{theorem}\label{thm4}
Suppose the hypotheses of Theorem \ref{thm3} are satisfied and that $b>0$ is
given. Then, for sufficiently small $T'>0$, there exist $\Gamma'>0$,
$\sigma\in(0,\,1)$, and $\sigma'\in(0,\,1)$, such that the approximation of
Theorem \ref{thm3} satisfies
$$
\left(\,\int_{|x-a(t)|>b}\ |\psi(x,t,\hbar)|^2\,dx\,\right)^{1/2}
\ \le\ \exp\,\left\{\,-\Gamma'/\hbar^{\sigma}\,\right\},
$$
whenever
$0\leq t\leq T'\ln(1/\hbar)$.\\
Moreover, if $\tau$ can be taken arbitrarily small, we can chose
$T'=\frac{1}{6 \lambda}(1-\eps)$ where $\eps$ is arbitrarily small.
\end{theorem}

We also explore the validity of the approximation in a scattering
framework and its consequences on the corresponding semiclassical approximation
of the scattering matrix $S(\hbar)$. This requires assumptions on the
decay of the potential and its derivatives at infinity.\\

For scattering theory, we assume $V$ satisfies the following decay hypothesis.\\

\noindent
{\bf D:}\quad {\em There exist $\beta>1$, $v_0>0$, and $v_1>0$, such that for
all $x\in\R^d$ and all multi-indices $m\in\N^d$,
\be\label{decay}
D^mV(x)\ \leq\ \frac{v_0\,v_1^{|m|}\,m!}{\langle x\rangle^{\beta+|m|}},
\ee
where $\langle x\rangle\,=\,\sqrt{1+x^2}$.}\\

Theorem 1.2 of \cite{scat} shows that under the hypothesis D,
the solution of the classical equations (\ref{newton})
satisfies the following asymptotic estimates:

For any $a_-\in\R^d$, $0\neq \eta_-\in\R^d$ such that
$(a_-,\,\eta_-)\in\R^{2d}\backslash{\cal E}$, where
${\cal E}\subseteq \{ (a_-,\,\eta_-)\in\R^{2d} : \eta_- \neq 0\}$ is
closed and of Lebesgue measure zero in $\R^d$, there exists
$(a_+,\,\eta_+)\in\R^{2d}$, $\eta_+\neq 0$, and $S_+\in\R$ such that
\bea\nonumber
& &\lim_{t\ra\pm\infty}\,|a(t)-a_{\pm}-\eta_{\pm}t|\ =\ 0,\\[4pt] \nonumber
& &\lim_{t\ra\pm\infty}\,|\eta(t)-\eta_{\pm}|\ =\ 0,\\[4pt] \nonumber
& &\lim_{t\ra -\infty}\,|S(t)-t\eta_-^2/2|\ =\ 0,\\[4pt]
& &\lim_{t\ra +\infty}\,|S(t)-S_+-t\eta_+^2/2|\ =\ 0.\label{clasflow}
\eea
Moreover, for any $d\times d$ matrices $(A_-,\,B_-)$ satisfying
condition (\ref{cond1}), there exist matrices $(A_+,\,B_+)\in M_d(\C)^2$
satisfying (\ref{cond1}), such that
\bea\nonumber
& &\lim_{t\ra\pm\infty}\,\| A(t)-A_{\pm}-iB_{\pm}t \|\ =\ 0, \\[4pt]
& &\lim_{t\ra\pm\infty}\,\| B(t)-B_{\pm}\|\ =\ 0.\label{linflow}
\eea

Our assumption D implies that $V$ is short range. It follows that if
$\ds H_0(\hbar)\ =\ -\,\frac{\hbar^2}{2}\,\Delta$, then the wave operators
defined by
\be\label{wavop}
\Omega^{\mp}(\hbar)\ =\ s-\lim_{s\ra \pm \infty}\,
\E^{iH(\hbar)s/\hbar}\,\E^{-iH_0(\hbar)s/\hbar}
\ee
exist and have identical ranges equal to the absolutely continuous subspace
of $H(\hbar)$. As a result, the scattering matrix
\be\label{sm}
S(\hbar)\ =\ \Omega^-(\hbar)^*\,\Omega^+(\hbar)
\ee
is unitary.

\vskip .5cm
\begin{theorem}\label{thm5}
Suppose $d\geq 3$ and assume hypothesis D. Let $(a_-,\,\eta_-)\in\R^{2d}
\backslash {\cal E}$ and\linebreak
$(A_-,\,B_-)\in M_d(\C)^2$ satisfy condition
(\ref{cond1}). Let $c_j(-\infty)\in\C$, for $j\in\N^d$, with $|j|\leq J$, such
that $\sum_{|j|\leq J}\,|c_j(-\infty)|^2=1$. Then, there exist
$(a_+,\,\eta_+)\in\R^{2d}$,  $(A_+,\,B_+)\in M_d(\C)^2$ satisfying
(\ref{cond1}),
$S_+\in\R$ and explicit coefficients $c_j(+\infty,\hbar)\in \C$,
for all $j\in\N^d$, with
$|j|\leq\tilde{J}_{\hbar}$ with $\tilde{J}_{\hbar}=J+3g/\hbar-3$
such that for some $\gamma>0$, $C>0$, $g>0$ (depending
on the the classical data), the states defined by
\bea\nonumber
\Phi_-(A_-, B_-, \hbar, a_-, \eta_-, x)&=&\sum_{|j|\leq J}c_j(-\infty)
\phi_j(A_-, B_-, \hbar, a_-, \eta_-, x)\\[5pt]\label{defastat}
\Phi_+(A_+, B_+, \hbar, a_+, \eta_+, x)&=&\E^{iS_+/\hbar}
\sum_{|j|\leq \tilde{J}_{\hbar}}c_j(+\infty,\hbar)
\phi_j(A_+, B_+, \hbar, a_+, \eta_+, x)
\eea
satisfy
$$
\|S(\hbar)\,\Phi_-(A_-, B_-, \hbar, a_-, \eta_-, \cdot)\,
-\,\Phi_+(A_+, B_+, \hbar, a_+, \eta_+, \cdot)\|_{L^2(\scriptsize{\R}^d)}
\ \leq\ C\,\E^{-\gamma/\hbar},
$$
if $\hbar$ is small enough.
\end{theorem}

Finally, we address the question of the generalization
of the initial coherent state, whose evolution
can be controlled up to exponential accuracy in the
different settings considered above.

For $(a,\eta)\in\R^{2d}$, we define $\Lambda_h(a,\eta)$ to be the operator
$$
(\Lambda_\hbar(a,\eta)f)(x)\,=\,
\hbar^{-d/2}\,\E^{i\bra\eta,\,(x-a)\ket/\hbar}\,f((x-a)/\sqrt{\hbar}).
$$
We define a dense set ${\cal C}$ in $L^2(\R^d)$, that is contained in the set
${\cal S}$ of Schwartz functions, by
\bea
{\cal C}&=&\left\{f(x)=\sum_{j}\,c_j\,\phi_j(\un,\,\un,\,1,\,0,\,0,\,x)\,
\in\,{\cal S},\right.\nonumber\\
&&\left. \mbox{ such that }\,\exists\, K>0 \mbox{ with }
\sum_{|j|>J}|c_j|^2\leq
\E^{-K J}, \mbox{ for large } J\right\}.\label{woofer}
\eea

\vskip .2cm \noindent
{\bf Remark}\quad It is easy to check that the inequality in (\ref{woofer}) is
equivalent to the requirement that the coefficients of $f$ satisfy
$$
|c_j|\,\leq\,\E^{-K|j|},
$$
for large $|j|$.
Another equivalent definition of ${\cal C}$ is
$$
{\cal C}\,=\,\cup_{t>0}\,\E^{-tH_{ho}}\,{\cal S},
$$
where $H_{ho}\,=\,-\,\Delta/2\,+\,x^2/2$ is the harmonic oscillator
Hamiltonian. The set ${\cal C}$ is sometimes called the set of analytic vectors
\cite{RS} for the harmonic oscillator Hamiltonian.

\vskip .5cm
\begin{theorem}\label{thm6}
All theorems above remain true if the initial condition
has the form
$$\psi(x,0,\hbar)=(\Lambda_\hbar(a,\eta)\ffi)(x),$$
where $\ffi\in {\cal C}$.
\end{theorem}

\vskip .5cm
Theorem \ref{thm1} is proved in Section \ref{truncsec}.
Theorem \ref{thm2} is proved in Section \ref{locsec}.
Theorems \ref{thm3} and \ref{thm4} are proved in Section \ref{longtime}.
Theorem \ref{thm5} is proved in Section \ref{scatt}.
Theorem \ref{thm6} is proved in Section \ref{initcond}.

\vskip .5cm
\section{An Alternative Semiclassical Expansion}
\setcounter{equation}{0}
\setcounter{theorem}{0}

\vskip .25cm
In this section we derive an expansion in powers of $\hbar^{1/2}$.
In later sections we perform optimal truncation of this expansion to obtain
exponentially accurate approximations.

\vskip .5cm
We wish approximately to solve the equation
\be\label{scheq}
i\,\hbar\,\frac{\partial\psi}{\partial t}\ = \
-\,\frac{\hbar^2}2\,\Delta\,\psi\,+\,V(x)\,\psi,
\ee
with initial conditions of the form
\be\label{init}
\psi(x,0,\hbar)\ =\
\sum_{|j|\le J}\,c_{0,j}(0)\,\phi_j(A(0),B(0),\hbar,a(0),\eta(0),x),
\ee
where $\ds\sum_{|j|\le J}\,\left|\,c_{0,j}(0)\,\right|^2\ =\ 1$.

We can write the exact solution to this equation in the basis of semiclassical
wave packets,
\be\label{expandpsi}
\psi(x,t,\hbar)\ =\ e^{iS(t)/\hbar}\
\sum_j\,c_j(t,\hbar)\,\phi_j(A(t),B(t),\hbar,a(t),\eta(t),x).
\ee
Note that the sum is over multi-indices $j$.
The infinite vector $c$ whose entries are the coefficients $c_j$ satisfies
\be\label{ceqn}
i\,\hbar\,\dot{c}\ =\ K(t,\hbar)\,c,
\ee
where $K(t,\hbar)$ is an infinite self--adjoint matrix.

The matrix $K(t,\hbar)$ has an asymptotic expansion in powers of
$\hbar^{1/2}$. The cubic term in the expansion of $V(x)$ around $x=a(t)$ gives
the leading non-zero term of order $\hbar^{3/2}$. The quartic term in the
expansion of $V(x)$ gives the term of order $\hbar^{4/2}$, etc.
Thus, we can write
\be\label{expandK}
K(t,\hbar)\ \sim\ \sum_{k=3}^\infty\,\hbar^{k/2}\,K_k(t),
\ee
with
\be\label{Kkdef}
K_k(t)\ = \ \sum_{|m|=k}\,\frac{(D^mV)(a(t))}{m!}\ X(t)^m,
\ee
where $X(t)^m$ is the infinite matrix that represents $\hbar^{-|m|/2}\,(x-a)^m$.
Explicit formulas \cite{raise} show that entries of $X(t)^m$ and $K_k(t)$ do not
depend on $\hbar$.

We formally expand the vector $c$ as
\bea\nonumber
c(t,\hbar)&=&
c_0(t)\,+\,\hbar^{1/2}\,c_1(t)\,+\,\hbar^{2/2}\,c_2(t)\,+\ \dots\\[5pt]
&=&\sum_k\,\hbar^{k/2}\,c_k(t).\label{expandc}
\eea
We denote the $j^{\mbox{\scriptsize th}}$ entry of $c_k(t)$ by $c_{k,j}(t)$.
Note that $k$ is a non-negative integer, and $j$ is a multi-index.
We substitute the two expansions (\ref{expandK}) and (\ref{expandc})
into (\ref{ceqn}) and divide by $\hbar$.
We then equate terms of the same orders on the two sides of the resulting
equation.

\vskip .5cm
\noindent {\bf Order $0$.}\quad The zeroth order terms simply require
\be\label{order0}
i\,\dot{c}_0\ =\ 0.
\ee
From (\ref{init}), the solution is obviously $c_{0,j}(t)\ =\ c_{0,j}(0)$.
We note that $c_{0,j}(t)=0$ if $|j|>J$.

\vskip .5cm
\noindent {\bf Order $1$.}\quad The first order terms require
\be\label{order1}
i\,\dot{c}_1\ =\ K_3(t)\,c_0(t).
\ee
We solve this by integrating. Because of (\ref{init}), $c_1(0)\,=\,0$.
From the form of $c_0(t)$, only finitely many of the entries of $c_1(t)$
are non-zero, and $c_{1,j}\,=\,0$ whenever $|j|>J+3$.
In $d$ space dimensions, $c_1(t)$ has at most
$\ds \left(\begin{array}{c} J+3+d\\ d\end{array}\right)$
non-zero entries.

\vskip .5cm
\noindent {\bf Order $2$.}\quad The second order terms require
\be\label{order2}
i\,\dot{c}_2\ =\ K_4(t)\,c_0(t)\,+\,K_3(t)\,c_1(t).
\ee
Again, we can solve this by integrating with $c_2(0)\,=\,0$.
The only entries of $c_2(t)$ that can be non-zero are $c_{2,j}(t)$
with $|j|\le J+6$. In $d$ dimensions, there are at most
$\ds \left(\begin{array}{c} J+6+d\\ d\end{array}\right)$
non-zero entries.

\vskip .5cm
\noindent {\bf Order $n$.}\quad In general, the $n^{\mbox{\scriptsize th}}$
order terms require
\be\label{ordern}
i\,\dot{c}_n\ =\ \sum_{k=0}^{n-1}\,K_{n+2-k}(t)\,c_k(t).
\ee
To solve this, we simply integrate. We observe that $c_{n,j}(t)$ can be non-zero
only if $|j|\le J+3n$. In $d$ dimensions, there are at most
$\ds \left(\begin{array}{c} J+3n+d\\ d\end{array}\right)$
non-zero entries.

\vskip .5cm
Our expansion is different from the one constructed in \cite{hagjoy3}, and it is
different from the Dyson expansion used in \cite{hagjoy3}. All three of these
expansions are asymptotic to the exact solution of the Schr\"odinger equation.
We note that the main construction in \cite{hagjoy3} yields a normalized wave
function. The expansion derived above does not generate normalized wavepackets.

\vskip .5cm
To prove that this expansion is asymptotic, we apply Lemma 2.8 of
\cite{raise}. To check the hypotheses of that lemma,
we do the expansion above through order $(l-1)$
to obtain $c_0(t)$, $c_1(t)$, \dots, $c_{l-1}(t)$. We substitute these into
(\ref{expandc}) with the sum cut off after $k=l-1$. We then use the result in
(\ref{expandpsi}) and compute
\be\label{truncation}
\xi_l(x,t,\hbar)\ =\
i\,\hbar\,\frac{\partial\psi}{\partial t}(x,t,\hbar)\ +\
\frac{\hbar^2}2\,\Delta\,\psi(x,t,\hbar)\,-\,V(x)\,\psi(x,t,\hbar)
\ee
Because the $c_k(t)$ solve (\ref{order0}), (\ref{order1}), (\ref{order2}), etc.,
there are many cancellations. We obtain
\bea\nonumber
\xi_l(x,t,\hbar)&=&e^{iS(t)/\hbar}\,
\sum_{k=0}^{l-1}\,\hbar^{k/2}\,W^{(l+1-k)}_{a(t)}(x)\,
\sum_{|j|\le\widetilde{J}(l)}\,
c_{k,j}(t)\,\phi_j(A(t),B(t),\hbar,a(t),\eta(t),x).
\\ && \label{xiexplicit}
\eea
Here, $\widetilde{J}(l)=J+3l-3$, and
for each $q$, $W^{(q)}_{a(t)}(x)$ denotes the Taylor series error
\bea\nonumber
W^{(q)}_{a(t)}(x)&=&
V(x)\,-\,\sum_{|m|\le q}\,\frac{D^mV(a(t))}{m!}\,(x-a(t))^m\\[5pt]
&=&\sum_{|m|=q+1}\,\frac{D^mV(\zeta_m(x,a(t))}{m!}\,(x-a(t))^m,\label{Wformula}
\eea
for some\ $\zeta_m(x,a(t))\,=\,a(t)+\theta_{m,x,a(t)}(x-a(t))$, with
$\theta_{m,x,a(t)}\in(0,\,1)$.

If $V$ is $C^{l+2}$ on some neighborhood of $\{ a(t)\,:\,t\in[0,\,T]\,\}$,
then each $W^{(q)}_{a(t)}(x)$ that occurs in (\ref{xiexplicit}) is bounded on a
slightly smaller neighborhood of $\{ a(t)\,:\,t\in[0,\,T]\,\}$. Since
$\left\|\,(x-a(t))^m\phi_j(A(t),B(t),\hbar,a(t),\eta(t),x)\,\right\|$ has order
$\hbar^{|m|/2}$, it follows that $\|\,\xi(\cdot,t,\hbar)\,\|$ has order
$\hbar^{l+2}$. Applying Lemma 2.8 of \cite{raise}, we learn that the
$\psi(x,t,\hbar)$
solves the Schr\"odinger equation up to an error whose norm is bounded by
$C_l\,\hbar^{l/2}$, when $\hbar$ is sufficiently small.

Note that the argument above requires the insertion of cutoffs to handle the
Gaussian tails or some other assumption, such as $V\in C^{l+2}(\R^d)$ with
$|\,D^mV(x)\,|\,\le\,M_m\,\exp(\tau x^2)$ for $|m|\le l+2$.

\vskip .5cm
\section{Estimates of the Expansion Coefficients}
\setcounter{equation}{0}
\setcounter{theorem}{0}

\vskip .25cm
In this section we study the behavior of $c_{k}(t)$.

\vskip .5cm
The first step is to get a good estimate of the operator norm of the bounded
operator $(x-a)^m\,P_{|j|\le n}$, where $P_{|j|\le n}$ denotes the projection
onto the span of the $\phi_j$ with $|j|\le n$.

\vskip .5cm
\begin{lemma}\label{opnorm}
In $d$ dimensions,
\be\label{simple}
(x-a)^m\,P_{|j|\le n}\,=\,P_{|j|\le n+|m|}\,(x-a)^m\,P_{|j|\le n},
\ee
and
\be\label{norm}
\|\,(x-a)^m\,P_{|j|\le n}\,\|\ \le \
\left(\,\sqrt{2\hbar}\ d\,\|\,A\,\|\,\right)^{|m|}\
\left(\,\frac{(n+|m|)!}{n!}\,\right)^{1/2}.
\ee
\end{lemma}

\vskip .5cm \noindent
{\bf Proof:}\quad
Formula (2.22) of \cite{raise} states that
$$(x_i-a_i)\ =\ \sqrt{\frac{\hbar}2}\
\left(\,\sum_p\,A_{i\,p}\,{\cal A}_p(A,B,\hbar,a,\eta)^*\,+\,
\sum_p\,\overline{A_{i\,p}}\,{\cal A}_p(A,B,\hbar,a,\eta)\,\right).$$
Note that the right hand side contains $2d$ terms.
Suppose $v$ is any vector in the range of $P_{|j|\le n}$.
Then using formulas (2.8) and (2.9) of \cite{raise}, we easily deduce that
\bea\nonumber
\|\,{\cal A}_p(A,B,\hbar,a,\eta)^*\,v\|&\le&\sqrt{n+1}\ \|\,v\,\|,\\[5pt]
\nonumber\|\,{\cal A}_p(A,B,\hbar,a,\eta)\,v\|&\le&\sqrt{n}\ \|\,v\,\|,
\eea
and that both ${\cal A}_p(A,B,\hbar,a,\eta)^*\,v$ and
${\cal A}_p(A,B,\hbar,a,\eta)\,v$ belong to the range of $P_{|j|\le n+1}$.

It follows immediately that
$$\|\,(x_i-a_i)\,P_{|j|\le n}\,\|\ \le \sqrt{2\hbar}\ d\,\|\,A\,\|\,\sqrt{n+1},
$$
and that $(x_i-a_i)\,P_{|j|\le n}\,=\,P_{|j|\le n+1}\,(x-a)\,P_{|j|\le n}$.

The lemma follows from these two results by a simple induction.\ep

\vskip .5cm
The conclusion to the next lemma contains the binomial coefficients
$\ds \left(\begin{array}{c} k-1\\ p-1\end{array}\right)$. For $k=1$ and $p=1$ we
define this to be $1$.

\vskip .5cm
\begin{lemma}\label{crucial} Suppose $V$ satisfies the hypotheses of Theorem
\ref{thm1}.

Fix $T$ and choose a classical orbit $a(t)$ for $0\le t\le T$. The hypotheses
guarantee that
\be\label{D1def}
D_1\ =\ \max\ \left\{\ 1,\ \sup_{0\le |n|,\ 0\le t\le T}\ \delta^{|n|}\
\frac{\left|\,(D^nV)(a(t))\,\right|}{n!}\ \right\}
\ee
and
\be\label{D2def}
D_2\ =\ \max\ \left\{\ 1,\ \sup_{\ 0\le t\le T}\
\sqrt{2}\ d\,\delta^{-1}\ \|\,A(t)\,\|\ \right\}
\ee
are finite.

We define $\ds D_3\ =\ \left(\begin{array}{c} d+2\\ d-1\end{array}\right)$,
which is the number of multi-indices $m$ with $|m|=3$ in $d$ dimensions.

Suppose $c_0(0)$ is a normalized vector with $c_{0,j}(0)$ non-zero only for
$|j|\le J$, and suppose $c_{k,j}(0)\,=\,0$ for all $j$ when $k\ge 1$.
Let $c_{k,j}(t)$ be the solution to (\ref{order0}), (\ref{order1}), \dots,
(\ref{ordern}), with these initial conditions.
Then for $t\,\in\,[0,\,T]$, we have
\be\label{sillyJ}
c_{0,j}(t)\ =\ 0\qquad \mbox{whenever}\qquad |j|>J,
\ee
\be\label{sillynorm}
\|\,c_0(t)\,\|\ \le\ D_1,
\ee
and for $k\ge 1$,
\be\label{yeah1}
c_k(t)\ =\ \sum_{p=1}^k\ c_k^{[p]}(t),
\ee
where
\be\label{yeah2}
c_{k,j}^{[p]}(t)\ =\ 0 \qquad \mbox{whenever}\qquad |j|>J+k+2p,
\ee
and
\be\label{yeah3}
\|\,c_k^{[p]}(t)\,\|\ \le\
\left(\begin{array}{c} k-1\\ p-1\end{array}\right)\
D_1^p\,D_2^{k+2p}\,D_3^k\
\left(\,\frac{(J+k+2p)!}{J!}\,\right)^{1/2}\ \frac{t^p}{p!}.
\ee
\end{lemma}

\vskip .5cm \noindent
{\bf Proof:}\quad The finiteness of $D_1$ and $D_2$ is standard.

The conclusions (\ref{sillyJ}) and (\ref{sillynorm}) are trivial.

We assume $t\in[0,\,T]$, and let $X(t)$ denote the formal vector whose
entries $X_i(t)$ denote the infinite matrix that represents
$\hbar^{-1/2}\,(x_i-a_i(t))$
in the basis $\{\,\phi_j(A(t),B(t),\hbar,a(t),\eta(t),\cdot)\,\}$.

From (\ref{order1}) we have
$$i\,\dot{c}_1(t)\ =\ K_3(t)\,c_0(t)\ =\ \sum_{|m|=3}\,
\frac{(D^mV)(a(t))}{m!}\,X(t)^m\,c_0(t).$$
We integrate to obtain $c_1(t)\,=\,c_1^{[1]}(t)$.
Lemma \ref{opnorm}, (\ref{sillyJ}), and (\ref{sillynorm})
imply two conclusions:
\be\label{c11}
c_{1,j}^{[1]}(t)\ =\ 0 \qquad \mbox{whenever}\qquad |j|>J+3,
\ee
and
\be\label{c11est}
\|\,c_1^{[1]}(t)\,\|\ \le\
D_1\,D_2^3\,D_3\ \left(\,\frac{(J+3)!}{J!}\,\right)^{1/2}\ t,
\ee
where the factor $\left(\begin{array}{c} d+2\\ d-1\end{array}\right)$ is the
number of multi-indices $m$ with $|m|=3$ in $d$ dimensions.
This proves (\ref{yeah1}), (\ref{yeah2}), and (\ref{yeah3}) for $k=1$.

For $k=2$, we have from (\ref{order2}),
\bea\nonumber
i\,\dot{c}_2(t)&=&K_4(t)\,c_0(t)\ +\ K_3(t)\,c_1(t)\\[4pt] \nonumber
&=&
\sum_{|m|=4}\,
\frac{(D^mV)(a(t))}{m!}\,X(t)^m\,c_0(t)\,+\,
\sum_{|m|=3}\,
\frac{(D^mV)(a(t))}{m!}\,X(t)^m\,c_1(t).
\eea
The two terms on the right hand side of this equation produce two terms,
$c_2^{[1]}(t)$ and $c_2^{[2]}(t)$, when we integrate to obtain $c_2(t)$.
Using (\ref{sillyJ}), (\ref{sillynorm}), (\ref{c11}), (\ref{c11est}), and two
applications of Lemma \ref{opnorm} we learn that
$c_2(t)=c_2^{[1]}(t)\,+\,c_2^{[2]}(t)$, where
\be\label{c21}
c_{2,j}^{[1]}(t)\ =\ 0 \qquad \mbox{whenever}\qquad |j|>J+4,
\ee
\be\label{c22}
c_{2,j}^{[2]}(t)\ =\ 0 \qquad \mbox{whenever}\qquad |j|>J+6,
\ee
\be\label{c21est}
\|\,c_2^{[1]}(t)\,\|\ \le\ \left(\begin{array}{c} d+3\\ d-1\end{array}\right)\
D_1\,D_2^4\,\left(\,\frac{(J+4)!}{J!}\,\right)^{1/2}\ t,
\ee
and
\be\label{c22est}
\|\,c_2^{[2]}(t)\,\|\ \le\
D_1^2\,D_2^6\,D_3^2\,\left(\,\frac{(J+6)!}{J!}\,\right)^{1/2}\ \frac{t^2}{2!}.
\ee
(The factor of $\ds \left(\begin{array}{c} d+3\\ d-1\end{array}\right)$ in
(\ref{c21est}) is the number of multi-indices $m$ with $|m|=4$.)
This implies (\ref{yeah1}), (\ref{yeah2}), and (\ref{yeah3}) for $k=2$ because
$\ds \left(\begin{array}{c} d+3\\ d-1\end{array}\right)\ \le\
\left(\begin{array}{c} d+2\\ d-1\end{array}\right)^2\ =\ D_3^2$. This
combinatorial inequality follows because $d\ge 1$ implies
\bea\nonumber
\left(\begin{array}{c} d+2\\ d-1\end{array}\right)^2\
\left(\begin{array}{c} d+3\\ d-1\end{array}\right)^{-1}&=&
\frac{(d+2)!^2\ 4!\ (d-1)!}{3!^2\ (d-1)!^2\ (d+3)!}\\
&=&\frac{4d}{d+3}\ \frac{d+2}3\ \frac{d+1}2
\ \ge\ \frac 4{1+3/d}\ \ge\ 1.\label{glorp}
\eea

From (\ref{ordern}) with $n=3$, we have
\bea\nonumber
i\,\dot{c}_3(t)&=&K_5(t)\,c_0(t)\,+\,K_4(t)\,c_1(t)\,+\,K_3(t)\,c_2(t)\\[5pt]
\nonumber&=&\sum_{k=0}^2\
\sum_{|m|=5-k}\,
\frac{(D^mV)(a(t))}{m!}\,X(t)^m\,c_k(t).
\eea
Using (\ref{sillyJ}), (\ref{sillynorm}), (\ref{c11}), (\ref{c11est}),
(\ref{c21}), (\ref{c22}), (\ref{c21est}), (\ref{c22est}),
and four applications of Lemma \ref{opnorm} we learn that
$c_3(t)=c_3^{[1]}(t)\,+\,c_3^{[2]}(t)\,+\,c_3^{[3]}(t)$, where
\be\label{c31}
c_{3,j}^{[1]}(t)\ =\ 0 \qquad \mbox{whenever}\qquad |j|>J+5,
\ee
\be\label{c32}
c_{3,j}^{[2]}(t)\ =\ 0 \qquad \mbox{whenever}\qquad |j|>J+7,
\ee
\be\label{c33}
c_{3,j}^{[3]}(t)\ =\ 0 \qquad \mbox{whenever}\qquad |j|>J+9,
\ee
\be\label{c31est}
\|\,c_3^{[1]}(t)\,\|\ \le\ \left(\begin{array}{c} d+4\\ d-1\end{array}\right)
D_1\,D_2^5\,\left(\,\frac{(J+5)!}{J!}\,\right)^{1/2}\ t,
\ee
\be\label{c32est}
\|\,c_3^{[2]}(t)\,\|\ \le\
\left(\begin{array}{c} d+3\\ d-1\end{array}\right)
2\,D_1^2\,D_2^7\,D_3\left(\,\frac{(J+7)!}{J!}\,\right)^{1/2}\ \frac{t^2}{2!},
\ee
and
\be\label{c33est}
\|\,c_3^{[3]}(t)\,\|\ \le\
D_1^3\,D_2^9\,D_3^3\left(\,\frac{(J+9)!}{J!}\,\right)^{1/2}\ \frac{t^3}{3!}.
\ee
This implies (\ref{yeah1}), (\ref{yeah2}), and (\ref{yeah3}) for $k=3$ because
of (\ref{glorp}) and the similar inequality
$$\left(\begin{array}{c} d+4\\ d-1\end{array}\right)\ \le\
\left(\begin{array}{c} d+2\\ d-1\end{array}\right)^3\ =\ D_3^3.$$
This inequality follows because $d\ge 1$ implies
$$
\left(\begin{array}{c} d+2\\ d-1\end{array}\right)^3\
\left(\begin{array}{c} d+4\\ d-1\end{array}\right)^{-1}\ =\
\frac{5}{1+4/d}\ \frac{4}{1+3/d}\ \left[\,\frac{d+2}3\ \frac{d+1}2\,\right]^2
\ \ge\ 1.
$$

Now suppose inductively that the lemma is true for all $k\le q$, for some
$q\ge 2$.
By integrating (\ref{ordern}) with $n=q+1$, we can decompose
\be\label{decomposec}
c_{q+1}(t)\ =\ c_{q+1}^{[1]}(t)\,+\,\sum_{n=1}^q\,\sum_{p=2}^{n+1}\,d[q,n,p](t),
\ee
where
\be\label{d0def}
c_{q+1}^{[1]}(t)\ =\ -\,i\,\int_0^t\,K_{q+3}(s)\,c_0(s)\,ds,
\ee
\be\label{dnpdef}
d[q,n,p](t)\ =\ -\,i\,\int_0^t\,K_{q+3-n}(s)\,c_n^{[p-1]}(s)\,ds,
\ee
for $1\le n\le q$ and $2\le p\le n+1$.
We interchange the sums in (\ref{decomposec}) to obtain
\be\label{yeah1q}
c_{q+1}(t)\ =\ c_{q+1}^{[1]}(t)\,+\,\sum_{p=2}^{q+1}\,c_{q+1}^{[p]}(t),
\ee
where
\be\label{cq1def}
c_{q+1}^{[p]}(t)\ =\ \sum_{n=p-1}^q\,d[q,n,p](t),
\ee
for $2\le p\le q+1$. This establishes (\ref{yeah1}) for $k=q+1$.

The induction hypotheses, formulas (\ref{Kkdef}), (\ref{d0def}), (\ref{dnpdef}),
(\ref{cq1def}),
and Lemma \ref{opnorm} imply (\ref{yeah2}) for $k=q+1$, as well as the two
inequalities
\be\label{cq1est}
\|\,c_{q+1}^{[1]}(t)\,\|\ \le\
\left(\begin{array}{c} d+q+2\\ d-1\end{array}\right)\
D_1\,D_2^{q+3}\,\left(\,\frac{(J+q+3)!}{J!}\,\right)^{1/2}\ t,
\ee
and
\bea\nonumber
\|\,d[q,n,p](t)\,\|&\le&
\left(\begin{array}{c} d+q+2-n\\ d-1\end{array}\right)\
D_1\,D_2^{q+3-n}\,\left(\,\frac{(J+q+1+2p)!}{(J+n+2p-2!}\,\right)^{1/2}\\[3pt]
\nonumber &&\quad\times\quad
\left(\begin{array}{c} n-1\\ p-2\end{array}\right)\
D_1^{p-1}\,D_2^{n+2p-2}\,D_3^n\
\left(\,\frac{(J+n+2p-2)!}{J!}\,\right)^{1/2}\ \frac{t^p}{p!}\\[7pt]
\nonumber &=&\left(\begin{array}{c} n-1\\ p-2\end{array}\right)\
\left(\begin{array}{c} d+q+2-n\\ d-1\end{array}\right)\
D_1^{p}\,D_2^{q+2p+1}\,D_3^n\\[7pt]
&&\qquad\qquad\times\quad
\left(\,\frac{(J+q+2p+1)!}{J!}\,\right)^{1/2}\ \frac{t^p}{p!}.\label{dnpest}
\eea
From these inequalities and (\ref{yeah1q}), we obtain (\ref{yeah3}) for $k=q+1$
as soon as we establish both the inequality
\be\label{Alptraum}
\left(\begin{array}{c} d+q+2-n\\ d-1\end{array}\right)\
\left(\begin{array}{c} d+2\\ d-1\end{array}\right)^n\ \le\
\left(\begin{array}{c} d+2\\ d-1\end{array}\right)^{q+1}
\ee
for $0\le n\le q$, and the identity
\be\label{binom}
\left(\begin{array}{c} q\\ p-1\end{array}\right)\ =\
\sum_{n=p-1}^q\,\left(\begin{array}{c} n-1\\ p-2\end{array}\right),
\ee
for $q\ge 2$ and $2\le p\le q+1$.

We set $r=q-n$ and note that (\ref{Alptraum}) is equivalent to
\be\label{Unsinn}
\left(\begin{array}{c} d+r+2\\ d-1\end{array}\right)\
\left(\begin{array}{c} d+2\\ d-1\end{array}\right)^{-r-1}\ \le\ 1,
\ee
for $0\le r\le q$.
However,
\bea\nonumber
&&\left(\begin{array}{c} d+r+2\\ d-1\end{array}\right)\
\left(\begin{array}{c} d+2\\ d-1\end{array}\right)^{-r-1}\\ \nonumber
&=&\frac{(d+2+r)!}{(d-1)!\ (r+3)!}\
\left(\,\frac{(d-1)!\ 3!}{(d+2)!}\,\right)^{r+1}\\ \nonumber
&=&\left[\,\frac{r+4}1\,\left(\frac 14\right)^{r+1}\,\right]\,
\left[\,\frac{r+5}2\,\left(\frac 25\right)^{r+1}\,\right]\,\cdots\,
\left[\,\frac{r+d+2}{d-1}\,\left(\frac{d-1}{d+2}\right)^{r+1}\,\right].
\eea
Inequality (\ref{Unsinn}) follows if each of the factors in the square brackets
is bounded by $1$. Thus, we need only prove
$$\frac{r+m+3}m\ \le \left(\,\frac{m+3}m\,\right)^{r+1},$$
for $1\le m$, which can be verified by using the binomial
expansion:
$$\frac{r+m+3}m\ =\ 1+\frac{r+3}m\ \le\ 1+(r+1)\frac 3m\,+\,\cdots\ =\
\left(1+\frac 3m\right)^{r+1}\ =\ \left(\,\frac{m+3}m\,\right)^{r+1}.
$$
This proves (\ref{Unsinn}) and hence (\ref{Alptraum}).

The identity (\ref{binom}) is trivial for $q=2$ and $p=2,\,3$. Assume
inductively
that it is true for all $2\le q\le m$ and $2\le p\le q+1$, where $m\ge 2$.
The identity (\ref{binom}) is trivial for $q=m+1$ and $p-1=m+1$, since
$$\left(\begin{array}{c} m+1\\ m+1\end{array}\right)\ =\ 1\ =\
\left(\begin{array}{c} m\\ m\end{array}\right).$$
Then for $m+1>p-1$, we have
\bea\nonumber
\left(\begin{array}{c} m+1\\ p-1\end{array}\right)&=&
\frac{(m+1)!}{(p-1)!\ (m-p+2)!}\\[4pt] \nonumber
&=&\frac{(p-1)\ (m!)}{(p-1)!\ (m-p+2)!}\ +\
\frac{(m-p+2)\ (m!)}{(p-1)!\ (m-p+2)!}\\[4pt] \nonumber
&=&\left(\begin{array}{c} m\\ p-2\end{array}\right)\ +\
\left(\begin{array}{c} m\\ p-1\end{array}\right)\\[4pt] \nonumber
&=&\left(\begin{array}{c} m\\ p-2\end{array}\right)\ +\
\sum_{n=p-1}^m\ \left(\begin{array}{c} n-1\\ p-2\end{array}\right)\\[4pt]
\nonumber
&=&\sum_{n=p-1}^{m+1}\ \left(\begin{array}{c} n-1\\ p-2\end{array}\right).
\eea
This proves (\ref{binom}) and completes the proof of the lemma.\ep

\vskip .5cm
\begin{corollary}\label{cor} Assume the hypotheses of Lemma \ref{crucial}.
Then in addition to (\ref{sillyJ}) and (\ref{sillynorm}), we have the following
for $k\ge 1$:
\be\label{yeah4}
c_{k,j}(t)\ =\ 0, \qquad \mbox{whenever}\qquad |j|>J+3k,
\ee
and
\be\label{yeah5}
\|\,c_k(t)\,\|\ \le\
\left(\,\frac{(J+3k)!}{J!}\,\right)^{1/2}\,\frac{D_2^k\,D_3^k}{k!}\,
\left(\,1\,+\,D_1\,D_2^2\,t\,\right)^{k}.
\ee
\end{corollary}

\vskip .5cm \noindent
{\bf Proof:}\quad Since $p\le k$, (\ref{yeah2}) implies (\ref{yeah4}).

To prove (\ref{yeah5}), we note that
$\ds\frac{((J+k+2p)!)^{1/2}}{p!}$ is increasing in $p$. Thus, (\ref{yeah3})
and $p\le k$ imply
$$
\|\,c_k^{[p]}(t)\,\|\ \le\
\left(\begin{array}{c} k-1\\ p-1\end{array}\right)\ D_1^p\ D_2^{k+2p}\ D_3^k\
\left(\,\frac{(J+3k)!}{J!}\,\right)^{1/2}\ \frac{t^p}{k!}.
$$
Summing over $p$, we obtain
$$
\|\,c_k(t)\,\|\ \le\
\left(\,\frac{(J+3k)!}{J!}\,\right)^{1/2}\ \frac{D_2^k\,D_3^k}{k!}
\left(\,1\,+\,D_1\,D_2^2\,t\,\right)^{k-1}\ D_1\,D_2^2\,t.
$$
This implies (\ref{yeah5}).\qquad\ep

\vskip .5cm
\section{Optimal Truncation Estimates}\label{truncsec}
\setcounter{equation}{0}
\setcounter{theorem}{0}

\vskip .5cm
In this section we show that the error given by (\ref{truncation}) and
(\ref{xiexplicit})
is exponentially small if we choose $l=\grintl\,g/\hbar\,\grintr$ for an
appropriate value of $g$.

The philosophy will be separately to estimate the error near the classical orbit
and far from the orbit. To do so, we
let $b$ be any positive number and define $\chi_1(x,t)$
to be the characteristic function of $\{\,x\,:\,|x-a(t)|\le b\,\}$. We set
$\chi_2(x,t)\,=\,1\,-\,\chi_1(x,t)$.

\vskip .5cm
\begin{lemma}\label{inside} Assume $V$ satisfies the hypotheses of Theorem
\ref{thm1}. Define $\chi_1(x,t)$ as above and $\xi_l(x,t,\hbar)$ by
(\ref{truncation}). For fixed $T>0$ and $b>0$,
there exists $G_1>0$, such that for each $g\in(0,\,G_1)$, there exist
$C_1$ and $\gamma_1>0$, such that
if $l$ is chosen to depend on $\hbar$ as
$l(\hbar)\,=\,\grintl\,g/\hbar\,\grintr$, then
\be\label{insideest}
\hbar^{-1}\
\int_0^T\,\|\,\chi_1(\cdot,t)\,\xi_{l(\hbar)}(\cdot,t,\hbar)\,\|\,dt\
\le\ C_1\,\exp\left\{\,-\gamma_1/\hbar\,\right\}.
\ee
\end{lemma}

\vskip .5cm \noindent
{\bf Proof:}\quad It is sufficient to prove the
existence of $\alpha_1$ and $\beta_1$, such that
\be\label{insideest1}
\hbar^{-1}\ \int_0^T\,\|\,\chi_1(\cdot,t)\,\xi_l(\cdot,t,\hbar)\,\|\,dt\ \le\
\alpha_1\,\beta_1^l\,l^{l/2}\,\hbar^{l/2}.
\ee
If this can be established, we choose $G_1=\beta_1^{-2}$. Then $0<g<G_1$ and
$l\,=\,\grintl\,g/\hbar\,\grintr$ imply $\beta_1^2\,g\,=\,e^{-\omega}$, with
$\omega>0$. Since
$\alpha_1\,(\beta_1^2\,l\,\hbar)^{l/2}\,=\,\alpha_1\,e^{-\omega g/(2\hbar)}$,
this implies the lemma with $C_1=\alpha_1$ and $\gamma_1=\omega g/2$.

To prove (\ref{insideest1}), we note first that our hypotheses imply the
finiteness of
\be\label{D4def}
D_4\ =\ \sup_{|n|\ge 0,\ 0\le t\le T,\ |x-a(t)|\le b}\ \delta^{|n|}\
\frac{\left|\,(D^nV)(x)\,\right|}{n!}.
\ee
We use this, (\ref{xiexplicit}), and (\ref{Wformula}) to see that
\bea\nonumber
&&\|\,\chi_1(x,t)\,\xi_l(x,t,\hbar)\,\|\\[8pt] \nonumber&\le&
\Bigg\|\,\sum_{k=0}^{l-1}\ \sum_{|m|=l+2-k}\,\hbar^{k/2}\,\,\chi_1(x,t)\,
\frac{(D^mV)(\zeta_m(x,a(t))}{m!}\,(x-a(t))^m\\
\nonumber &&\qquad\qquad\qquad\qquad\qquad\qquad\qquad\times\quad
\sum_{|j|\le\widetilde{J}(l)}\,c_{k,j}(t)\,
\phi_j(A(t),B(t),\hbar,a(t),\eta(t),x)\,\Biggr\|\\[6pt]\nonumber
&\le&\sum_{k=0}^{l-1}\,\hbar^{k/2}\,D_4\,\delta^{-l-2+k}\\ \nonumber
&&\qquad\qquad\times\quad\sum_{|m|=l+2-k}\,\Bigg\|\,(x-a(t))^m\,
\sum_{|j|\le\widetilde{J}(l)}\,c_{k,j}(t)\,
\phi_j(A(t),B(t),\hbar,a(t),\eta(t),x)\,\Bigg\| \\[6pt]
&\le&D_4\ \hbar^{l/2+1}\,\sum_{k=0}^{l-1}\ \delta^{-l-2+k}\,\sum_{|m|=l+2-k}\,
\left\|\,X(t)^m\,c_{k}(t)\,\right\|,\nonumber
\eea
where $X(t)$ is the infinite matrix that represents
$\hbar^{-1/2}(x-a(t))$ in the $\phi_j$ basis.

Thus,
\bea\nonumber
\hbar^{-1}\ \int_0^T\,\|\,\chi_1(\cdot,t)\,\xi_l(\cdot,t,\hbar)\,\|\,dt&\le&
D_4\,\hbar^{l/2}\,\sum_{k=0}^{l-1}\,
\int_0^T\,\delta^{-l-2+k}\,\sum_{|m|=l+2-k}\,
\left\|\,X(t)^m\,c_{k}(t)\,\right\|\,dt.\\[4pt]&&\label{dog}
\eea

We apply Lemmas \ref{opnorm} and \ref{crucial} to estimate each integral
on the right hand side of (\ref{dog}). For $k=0$, we obtain
\bea\nonumber
\int_0^T\,\delta^{-l-2}\,\sum_{|m|=l+2}\,
\left\|\,X(t)^m\,c_{0}(t)\,\right\|\,dt
&\le&D_1\ D_2^{l+2}\ \left(\begin{array}{c} d+l+1\\ d-1\end{array}\right)\,
\left(\,\frac{(J+l+2)!}{J!}\,\right)^{1/2}\ T\\
&\le&\left(\,\frac{(J+3l)!}{J!}\,\right)^{1/2}\
\frac{D_2^{l+2}\,D_3^l}{(l-1)!}\ D_1\ D_2^2\ T.\label{dog0}
\eea
In the last step, we have used $D_2\ge 1$, (\ref{Alptraum}), and
$\ds (J+l+2)!\,\le\,\frac{(J+3l)!}{((l-1)!)^2}$, which is true for $l\ge 1$.

For $k\ge 1$, we write the integral on the right hand side of (\ref{dog})
as a sum of $k$ terms by employing (\ref{yeah1}).
By (\ref{yeah2}), (\ref{yeah3}), and Lemma \ref{opnorm}, the
$p^{\mbox{\scriptsize th}}$ integrand satisfies
\bea\nonumber
&&\delta^{-l-2+k}\,\sum_{|m|=l+2-k}\,
\left\|\,X(t)^m\,c_{k}^{[p]}(t)\,\right\|\\ \nonumber
&\le&D_2^{l+2-k}\ \left(\begin{array}{c} d+l+1-k\\ d-1\end{array}\right)\
\left(\,\frac{(J+l+2p+2)!}{(J+k+2p)!}\,\right)^{1/2}
\\[4pt] \nonumber &&\quad\qquad\qquad\qquad\times\quad
\left(\begin{array}{c} k-1\\ p-1\end{array}\right)\,
D_1^p\ D_2^{k+2p}\,D_3^k\,\left(\,\frac{(J+k+2p)!}{J!}\,\right)^{1/2}\,
\frac{t^p}{p!}\\[6pt] \nonumber
&=&\left(\begin{array}{c} k-1\\ p-1\end{array}\right)\,
D_1^p\ D_2^{l+2p+2}\,D_3^k\,
\left(\begin{array}{c} d+l+1-k\\ d-1\end{array}\right)\,
\left(\,\frac{(J+l+2p+2)!}{J!}\,\right)^{1/2}\,
\frac{t^p}{p!}\\[5pt]
&\le&\left(\begin{array}{c} k-1\\ p-1\end{array}\right)\,
D_1^p\ D_2^{l+2p+2}\,D_3^l\,
\left(\,\frac{(J+l+2p+2)!}{J!}\,\right)^{1/2}\,
\frac{t^p}{p!}.\label{doga}
\eea
In the last step, we have again used (\ref{Alptraum}).

We now mimic the proof of Corollary \ref{cor} and then integrate to obtain
%
%
the following estimate of the $k^{\mbox{\scriptsize th}}$
term in (\ref{dog}):
$$
\left(\,\frac{(J+3l)!}{J!}\,\right)^{1/2}\,\frac{D_2^{l+2}}{(l-1)!}\
\frac{D_3^l}{(k+1)D_1\,D_2^2}\,
\left[\,\left(\,1\,+\,D_1\,D_2^2\,T\,\right)^{k+1}\,-\,1\,\right].
$$

We define
\be\label{D5def}
D_5\ =\ 1\,+\,D_1\,D_2^2\,T,
\ee
bound $\ds\frac{D_3^l}{(k+1)D_1\,D_2^2}$ by $D_3^l$, and
$\left[\,\left(\,1\,+\,D_1\,D_2^2\,T\,\right)^{k+1}-\,1\,\right]$ by
$D_5^{k+1}$. We then sum over $k$ in (\ref{dog}) to obtain the estimate
\be\label{explicitJ}
\hbar^{-1}\ \int_0^T\,\|\,\chi_1(\cdot,t)\,\xi_l(\cdot,t,\hbar)\,\|\,dt
\ \le\ D_4\,\hbar^{l/2}\,
\left(\,\frac{(J+3l)!}{J!}\,\right)^{1/2}\
\frac{D_2^{l+2}\,D_3^l}{(l-1)!}\ \frac{D_5^{l+1}}{D_5-1}.
\ee
In this expression we bound $(J+3l)!$ by $(J+3l)^{J+3l}$ and
$\ds\frac{1}{(l-1)!}$ by $\ds\frac{1}{\rho^{l-1}\,(l-1)^{l-1}}$, which holds for
some constant $\rho$.
After some algebra, this leads to the estimate (\ref{insideest1}). \ep

\vskip .5cm
To prove the analogous lemma with $\chi_1$ replaced by $\chi_2$, we use
spherical coordinates for $\hbar^{-1/2}(x-a)$.

In spherical coordinates when $d\ge 2$, the operator $-\Delta_y\,+\,y^2$ has the
form
$$-\,\frac{\partial^2\phantom{r}}{\partial r^2}\ -\
\frac{d-1}r\,\frac{\partial\phantom{r}}{\partial r}\ +\
\frac{{\cal L}^2}{r^2}\ +\ r^2.$$
Here ${\cal L}^2$ is the Laplace--Beltrami operator on $S^{d-1}$. For $d\ge 3$,
it has eigenvalues
\be\label{sphevs}
\lambda_q\ =\ q(q+d-2),
\ee
with multiplicities
\bea\nonumber
m_q&=&\frac 1{(d-1)!}\,(q+1)\,(q+2)\,\cdots\,(q+d-3)\,
\,\bigg\{\,(q+d-2)(q+d-1)-(q-1)q\,\bigg\}\\ \nonumber
&=&\frac 1{(d-2)!}\,(q+1)\,(q+2)\,\cdots\,(q+d-3)\ \,
(d-1)(2q+d-2)\\[4pt]
&\le&C_d\ e^{\alpha_d q},\label{sphmult}
\eea
where $q\,=\,0,\,1,\,\dots$. We denote a corresponding orthonormal basis of
eigenfunctions by $Y_{q,m}(\omega)$ for $1\le m\le m_q$.

When $d=1$, the analog of the Laplace--Beltrami operator is multiplication by
$\lambda=0$ on even functions and multiplication by $\lambda=1$ on odd
functions. The operator
$-\,\frac{\partial^2\phantom{r}}{\partial x^2}\,+\,x^2$ on $\R$
just becomes the direct sum of two copies of
$-\,\frac{\partial^2\phantom{r}}{\partial r^2}\,+\,r^2$
on $(0,\,\infty)$ with Neumann and Dirichlet boundary conditions at $r=0$.

When $d=2$, (\ref{sphmult}) should be replaced with $m_0=1$ and $m_q=2$ for
$q>0$, but the inequality $m_q\,\le\,C_d\,\E^{a_dq}$ still holds.

The eigenvalues of
$-\Delta_y\,+\,y^2$ are $E=4n+2q+d$ with normalized eigenfunctions
\be\label{radstates}
\psi_{q,n,m}(r,\omega)\ =\ \sqrt{\frac{2n!}{\Gamma(q+n+\frac d2)}}\ \,
r^q\ L_n^{q+\frac d2-1}(r^2)\ \ e^{-r^2/2}\ Y_{q,m}(\omega).
\ee
Here
\be\label{laguerre} L_n^{\beta}(x)\ =\ \sum_{m=0}^n\ (-1)^m\
\left(\begin{array}{c} n+\beta\\ n-m\end{array}\right)\ \frac{x^m}{m!}
\ee
denotes the Laguerre polynomial that satisfies the differential equation
\be\label{laguerreODE}
x\,u''(x)\ +\ (\beta-x+1)\,u'(x)\ +\ n\,u(x)\ =\ 0,
\ee
and the normalization condition
\be\label{normalized}
\int_0^{\infty}\ L_n^{\beta}(x)\ L_m^{\beta}(x)\ x^{\beta}\ e^{-x}\ dx\ =\
\left\{\,\begin{array}{cl} 0&{\mbox{if }}n\ne m\\
\frac{\Gamma(\beta+n+1)}{n!}&{\mbox{if }}n=m\end{array}\right.,
\ee
for $\beta\,>\,-1$.

\vskip .5cm
The following lemma implies an estimate for $|\psi_{q,n,m}(r,\omega)|$ when $r$
is in the region which is classically forbidden because of energy
considerations.

\vskip .5cm
\begin{lemma}\label{db-est} For $\beta=q+\frac d2-1$ with $q=0,\,1,\,\dots$, the
Laguerre polynomial $\ds L_n^{\beta}(x)$ in (\ref{radstates}) satisfies
$\ds |L_n^{\beta}(x)|\ \le\ \frac{x^n}{n!}$ whenever
$x\ >\ 4n+2\beta+2\,=\,4n+2q+d$.
\end{lemma}

\vskip .5cm \noindent
{\bf Proof:}\quad
We mimic the proof of Lemma 3.1 of \cite{hagjoy3}. The first step is to show
that $g(r)\ =\ r^{\beta}\ L_n^{\beta}(r^2)\ \ e^{-r^2/2}$ cannot
vanish in the classically forbidden region $r^2\,>\,4n+2q+d$. This function
vanishes at infinity and is a non-trivial solution to an equation of the form
$$-\,g''(r)\ +\ w(r)\,g(r)\ =\ 0,$$
where $w(r)>0$ for $r^2\,>\,4n+2q+d$.
From this differential equation we conclude that $g$ and
$g''$ have the same sign in this region. By standard uniqueness theorems, $g$
and $g'$ cannot both vanish at the same point. To obtain a contradition, suppose
$g$ has a zero at some point $r_1$ with $r_1^2\,>\,4n+2q+d$.
Since $g$ vanishes at infinity, the mean value theorem guarantees
that $g'(r_2)=0$ for some $r_2>r_1$. Without loss of generality, we may assume
$g(r_2)>0$. This forces $g''(r_2)>0$, so $g'$ is locally increasing. It follows
that $g'$ is increasing for all $r>r_2$. Thus,
$g$ could not go to zero at infinity. This contradiction shows that $g$ could
not have had a zero in the region.

We now proceed by induction on $n$. Since $L^{\beta}_0(x)\,=\,1$, the lemma is
true for $n=0$. We now assume $n\,\ge\,1$ and that
the lemma has been established for $L^{\beta}_{n-1}(x)$.

Our non-vanishing result and (\ref{laguerre}) imply
$$\frac{L^{\beta}_n(x)\ n!}{(-1)^n\ x^n}\ =\ 1\ -\ B_{\beta,\,n}(x),$$
where $B_{\beta,\,n}(x)\,=\,O(1/x)$ for large $x$, and
$B_{\beta,\,n}(x)\,>\,-1$, for $x\,>\,4n+2\beta+2$.

Using recurrence relation 8.971.3 of \cite{GR}, we have
\bea\nonumber
\frac{d\phantom{x}}{dx}\,B_{\beta,\,n}(x)
&=&\frac{x\,{L_n^{\beta}}'(x)\,-\,n\,L^{\beta}_n(x)}{x^{n+1}}\
\frac{n!}{(-1)^n}\\ \nonumber
&=&-\ \frac{(n+\beta)\,L^{\beta}_{n-1}(x)}{x^{n+1}}\ \frac{n!}{(-1)^n}.
\eea
By our induction hypothesis, $L^{\beta}_{n-1}(x)$ has sign $(-1)^{n-1}$ for
$x\,>\,4n+2\beta-2$, which includes the region of interest.

Thus, $B_{\beta,\,n}(x)$ is increasing. Since it goes to zero at infinity, it
cannot be positive. This implies the lemma.\qquad\ep

\vskip .5cm
\begin{lemma}\label{outside} Assume $V$ satisfies the hypotheses of Theorem
\ref{thm1}. Define $\chi_2(x,t)$ as above and $\xi_l(x,t,\hbar)$ by
(\ref{truncation}).
%
For fixed $T>0$ and $b>0$,
there exists $G_2>0$, such that for each $g\in(0,\,G_2)$, there exist
$C_2$ and $\gamma_2>0$, such that
if $l$ is chosen to depend on $\hbar$ as
$l(\hbar)\,=\,\grintl\,g/\hbar\,\grintr$, and $\hbar$ is sufficiently small,
then
\be\label{outsideest}
\hbar^{-1}\
\int_0^T\,\|\,\chi_2(\cdot,t)\,\xi_{l(\hbar)}(\cdot,t,\hbar)\,\|\,dt\ \le\
C_2\,\exp\left\{\,-\gamma_2/\hbar\,\right\}.
\ee
\end{lemma}

\vskip .5cm \noindent
{\bf Proof:}\quad
We begin by using the analyticity of $V$ to control Taylor series errors.
We define
$$
C_{\delta}(x)=\{z\in\C^d\,:\ z_j=x_j+\delta e^{i\theta_j},\,
\theta_j\in [0,2\pi),\,j=1,2,\cdots, d\} .
$$
If $z\in C_{\delta}(\zeta(x,a))$, then, for all
$j=1,2,\cdots ,d$,
$$
|z_j|\,\leq\,\delta +|\zeta_j(x,a)|\,\leq\,\delta +|a_j|+|x_j-a_j|.
$$
Using this and applying $(b+c)^2\leq 2(b^2+c^2)$ several times,
we see that
$z\in C_{\delta}(\zeta(x,a))$ implies
$$
|\,V(z)\,|\,\leq\,M\,\exp(2\tau(x-a)^2)\,\exp(4\tau(\delta^2 d+a^2)).
$$
Hence, writing $\ds \frac{1}{m!}\,D^mV(\zeta(x,a))$ as a $d$--dimensional Cauchy
integral, we obtain the bound
$$
\frac{1}{p!}\ |D^pV(\zeta(x,a))|\ \leq\
M\ \frac{\exp(4\tau(\delta^2 d+a^2))}{\delta^{|p|}}\
\exp(2\tau(x-a)^2),
$$
where $\zeta(x,a)$ is any value between $x$ and $a$. Thus, for $0\le t\le T$,
there exists a constant $M_1$, such that
\be\label{mess8}
\frac{1}{p!}\ |(D^pV)(\zeta_p(x,a(t)))|\ \le\
\frac{M_1}{\delta^{|p|}}\ \exp(2\tau(x-a(t))^2).
\ee

We use this, (\ref{xiexplicit}), (\ref{Wformula}), and (\ref{yeah4})
to see that
\bea\nonumber
&&\|\,\chi_2(x,t)\,\xi_l(x,t,\hbar)\,\|\\[6pt] \nonumber&\le&
\Bigg\|\,\sum_{k=0}^{l-1}\,\hbar^{k/2}\,\,\chi_2(x,t)\,
\sum_{|p|=l+2-k}\,
\frac{(D^pV)(\zeta_p(x,a(t))}{p!}\,(x-a(t))^p\\
\nonumber &&\qquad\qquad\qquad\qquad\qquad\qquad\quad\ \times\quad
\sum_{|j|\le J+3k}\,c_{k,j}(t)\,
\phi_j(A(t),B(t),\hbar,a(t),\eta(t),x)\,\Biggr\|
\\[6pt]\nonumber
&\le&\sum_{k=0}^{l-1}\,\hbar^{k/2}\,
M_1\,\delta^{-l-2+k}\,\sum_{|p|=l+2-k}\,
\Bigg\|\,\chi_2(x,t)\,\exp(2\tau(x-a(t))^2)\,(x-a(t))^p\\
\nonumber &&\qquad\qquad\qquad\qquad\qquad\qquad\quad\ \times\quad
\sum_{|j|\le J+3k}\,c_{k,j}(t)\,
\phi_j(A(t),B(t),\hbar,a(t),\eta(t),x)\,\Bigg\|
\\[6pt]\nonumber
&\le&M_1\,\delta^{-l-2}\sum_{k=0}^{l-1}\,\hbar^{k/2}\,
\delta^{k}\sum_{|j|\le J+3k}\,|c_{k,j}(t)|\,\sum_{|p|=l+2-k}\,
\Bigg\|\,\chi_2(x,t)\,e^{2\tau(x-a(t))^2}\,(x-a(t))^p\\
\nonumber
&&\qquad\qquad\qquad\qquad\qquad\qquad\qquad\qquad\qquad\quad\times\quad\
\phi_j(A(t),B(t),\hbar,a(t),\eta(t),x)\,\Bigg\|.
\\[5pt]&&\label{awful}
\eea
Note that (\ref{yeah4}) has been used to replace $|j|\le\widetilde{J}(l)$
with $|j|\le J+3k$.

The norm in the final expression of (\ref{awful}) equals
\be\label{target}
\left\|\,\chi_{\{\,z:\,|z|>b\,\}}(x)\,\exp(2\tau x^2)\,x^p\,
\phi_j(A(t),B(t),\hbar,0,0,x)\,\right\|,
\ee
where $|p|=l+2-k$.

We assume that $\hbar$ is sufficiently small that $4\tau\hbar|A(t)|^2\,<\,2/3$.
Then the square of the quantity (\ref{target}) equals
\bea\nonumber
&&2^{-|j|}\,(j!)^{-1}\,\pi^{-d/2}\,|\det\,A(t)|^{-1}\,\hbar^{-d/2}
\\[5pt]\nonumber
&&\qquad\qquad\qquad\qquad\times\quad\int_{|x|>b}x^{2p}e^{4\tau x^2}\,
|{\cal H}_j(A;\,|A(t)|^{-1}\,\hbar^{-1/2}x)|^2\,
e^{-|A(t)|^{-2}x^2/\hbar}\,dx\\[7pt]\nonumber
&\le&
\frac{\left(\hbar\,\|A(t)\|^2\right)^{|p|}}{2^{|j|}\,(j!)\,\pi^{d/2}}\
\int_{\hbar^{1/2}|\,|A(t)|\,y\,|>b}\,
|y|^{2|p|}\,|{\cal H}_j(A;\,y)|^2\,e^{(4\tau\hbar|A(t)|^2-1)y^2}\,dy.\\[7pt]
\nonumber &\le&
\frac{\left(\hbar\,\|A(t)\|^2\right)^{|p|}}{2^{|j|}\,(j!)\,\pi^{d/2}}\
\int_{\hbar^{1/2}|\,|A(t)|\,y\,|>b}\,
|y|^{2|p|}\,|{\cal H}_j(A;\,y)|^2\,e^{-y^2 2/3}\,dy\\[7pt]
&\le&
\E^{-b^2/(6\|A(t)\|^2\hbar)}\frac{\left(\hbar\,\|A(t)\|^2\right)^{|p|}}
{2^{|j|}\,(j!)\,\pi^{d/2}}\
\int_{\hbar^{1/2}|\,|A(t)|\,y\,|>b}\,
|y|^{2|p|}\,|{\cal H}_j(A;\,y)|^2\,e^{-y^2 1/2}\,dy.
\label{yuk1}
\eea

By formula (3.7) of \cite{hagjoy3},
$\ds\Omega_j(y)\ =\ \sqrt{\frac{1}{2^{|j|}\,j!\,\pi^{d/2}}}\
{\cal H}_j(A;\,y)\
e^{-y^2/2}$ is a normalized eigenfunction of $-\Delta_y\,+\,y^2$ with eigenvalue
$2|j|+d$.
Thus, in spherical coordinates, it can be written as
\be\label{chgbasis}
\Omega_j(y)\ =\ \sum_{\{q,n,m:\,2n+q=|j|\}} d_{j,q,n,m}\,\psi_{q,n,m}(r,\omega),
\ee
where $\ds\sum_{\{q,n,m:\,2n+q=|j|\}} |d_{j,q,n,m}|^2\ =\ 1$.

We ultimately choose $l\,=\,\grintl\,g/\hbar\,\grintr$, with $0<g<G_2$. Since
$\widetilde{J}(l)\,=\,J+3l-3$, there exists $C_3$, such that $\hbar<1$ implies
$\widetilde{J}(l)\le C_3/\hbar$. By choosing $G_2$ sufficiently small, we
also have $\widetilde{J}(l)\,<\,(\|A(t)\|^{-2}b^2-1)/(2\hbar)$ for $0\le t\le T$
and small $\hbar$.
Thus, the relevant values of $j$ in (\ref{yuk1}) satisfy
$\sqrt{2|j|+d}\,<\,\|A(t)\|^{-1}b\hbar^{-1/2}$.

Lemma \ref{db-est} shows that
$\ds \left|\,r^q\,L_n^{q+\frac d2-1}(r^2)\,\right|\ \le\ \frac{r^{2n+q}}{n!}$
whenever $r^2\ >\ 4n+2q+d=2|j|+d$.
So, we see that (\ref{yuk1}) is bounded by
$$
\hbar^{|p|}\,\|A(t)\|^{2|p|}\,\int_{S^{d-1}}
\int_{r>\frac{b}{\|A\|\hbar^{1/2}}}\ r^{2|p|}\,
\left|\,\sum_{\{q,n,m:\,2n+q=|j|\}}\!\!
d_{j,q,n,m}\psi_{q,n,m}(r,\omega)\,\right|^2\,e^{r^2/2}\,r^{d-1}dr\,d\omega.
$$
We interchange the sum and integrals and apply the Schwartz inequality to the
sum. This shows that (\ref{yuk1}) is bounded by
\bea\nonumber
&&\E^{-b^2/(6\|A(t)\|^2\hbar)}\,\hbar^{|p|}\,\|A(t)\|^{2|p|}\,
\sum_{\{q,n,m:\,2n+q=|j|\}}\,\frac{2 n!}{\Gamma(q+\frac d2+n)}\\[4pt]\nonumber
&&\qquad\quad\qquad\times\quad
\int_{\frac{b}{\|A\|\hbar^{1/2}}}^\infty r^{d-1+2|p|+2q}\,
\left|\,L_n^{q+\frac d2-1}(r^2)\,\right|^2\,e^{-r^2/2}\,dr\,
\int_{S^{d-1}}\,\left|\,Y_{q,m}(\omega)\,\right|^2\,d\omega\\[7pt]\nonumber
&=&\E^{-b^2/(6\|A(t)\|^2\hbar)}\,
\hbar^{|p|}\,\|A(t)\|^{2|p|}\,\sum_{\{q,n,m:\,2n+q=|j|\}}\,
\frac{2 n!}{\Gamma(q+\frac d2+n)}\,\\[7pt]\nonumber
&&\qquad\quad\qquad\qquad\qquad\times\quad
\int_{\frac{b}{\|A\|\hbar^{1/2}}}^\infty r^{d-1+2|p|+2q}
\left|L_n^{q+\frac d2-1}(r^2)\right|^2e^{-r^2/2}\,dr.\\[5pt]
&&\label{page7}
\eea
By reducing the value of $G_2$ if necessary, we can ensure that the hypotheses
of Lemma \ref{db-est} are satisfied in the integration region in the right hand
side of (\ref{page7}). So, Lemma \ref{db-est} shows that the integral
satisfies
\bea\nonumber
&&\int_{\frac{b}{\|A\|\hbar^{1/2}}}^\infty\,r^{d-1+2|p|+2q}\,
\left|\,L_n^{q+\frac d2-1}(r^2)\,\right|^2\,e^{-r^2/2}\,dr\\[5pt]\nonumber
&\le&\frac{1}{(n!)^2}\,
\int_{\frac{b}{\|A\|\hbar^{1/2}}}^\infty\,r^{4n+d-1+2|p|+2q}\,e^{-r^2/2}\,dr
\\[5pt]\nonumber
&\le&\frac{2^{2n+\frac d2+q+|p|}}{(n!)^2}\,
\int_0^\infty\,z^{4n+d-1+2q+2|p|}\,e^{-z^2}\,dz
\\[5pt]\nonumber
&=&\frac{2^{2n+\frac d2+q+|p|-1}}{(n!)^2}\,
\Gamma(2n+\frac d2+q+|p|).
\eea
So, (\ref{page7}) is bounded by
$$\E^{-b^2/(6\|A(t)\|^2\hbar)}\hbar^{|p|}\,\|A(t)\|^{2|p|}\,\sum_{\{q,n,m:\,2n+q=|j|\}}\
\frac{2^{2n+\frac d2+q+|p|}}{n!}\
\frac{\Gamma(2n+\frac d2+q+|p|)}{\Gamma(q+\frac d2+n)}.$$
We use (\ref{sphmult}) to estimate the sum over $m\le m_q$ and bound this by
\bea\nonumber
&&C_d\,\E^{-b^2/(6\|A(t)\|^2\hbar)}\,\hbar^{|p|}\,\|A(t)\|^{2|p|}\,
\sum_{\{n,q:\,q=|j|-2n\}}\,
e^{\alpha_d q}\ \frac{2^{|j|+\frac d2+|p|}}{n!}\
\frac{\Gamma(|j|+\frac d2+|p|)}{\Gamma(q+\frac d2+n)}\\[4pt]\nonumber
&=&
C_d\E^{-b^2/(6\|A(t)\|^2\hbar)}\,\hbar^{|p|}\,\|A(t)\|^{2|p|}\,2^{|j|+
\frac d2+|p|}\,e^{\alpha_d|j|}\,
\Gamma(|j|+\frac d2+|p|)\,\\[4pt]\nonumber
&&\qquad\qquad\qquad\qquad\qquad\qquad\qquad\quad\qquad\times\quad
\sum_{n\le |j|/2}\
\frac{e^{-2\alpha_dn}}{n!\ \Gamma(q+\frac d2+n)}.
\eea
Since $e^{-2\alpha_dn}\le 1$, this is bounded by
\bea\nonumber
C_d'\,\E^{-b^2/(6\|A(t)\|^2\hbar)}\,\hbar^{|p|}\,
\|A(t)\|^{2|p|}\,2^{|j|+|p|}\,e^{\alpha_d|j|}\,
\Gamma(|j|+\frac d2+|p|)\,
\sum_{n\le |j|/2}\,
\frac{1}{n!\,\Gamma(q+\frac d2+n)}.\\[4pt] \label{berlin1}
\eea
For $n\le |j|/2$ and $d$ fixed, there exists a constant $C''$, such that
$\ds \Gamma(|j|-n+\frac d2)\,\ge\,C''\,(|j|-n)!$. So,
the sum over $n$ in (\ref{berlin1}) is bounded by
$$
\frac 1{C''}\ \sum_{n\le |j|/2}\ \frac 1{n!\,(|j|-n)!}\ \le\
\frac 1{C''\,|j|!}\ \sum_{n\le |j|/2}\
\left(\begin{array}{c}|j|\\n\end{array}\right)\
\ \le\ C'''\ \frac{2^{|j|}}{|j|!}.
$$
Thus, (\ref{page7}) is bounded by
\be\label{berlin2}
C''''\, \E^{-b^2/(6\|A(t)\|^2\hbar)}\ \hbar^{|p|}\ \|A(t)\|^{2|p|}\ 2^{|p|}\
e^{\beta_d|j|}\ \,\frac{\Gamma(|j|+\frac d2+|p|)}{|j|!}.
\ee
This quantity bounds (\ref{yuk1}), which, in turn, bounds the square of
(\ref{target}). Terms of the form (\ref{target}) occur in (\ref{awful}).
Putting this all together, we see that (\ref{awful}) is bounded by
\bea\nonumber
&&M_1\,{C''''}^{1/2}\,\E^{-b^2/(12\|A(t)\|^2\hbar)}\,\delta^{-l-2}\,
\sum_{k=0}^{l-1}\,\hbar^{k/2}\,\delta^k\,
\sum_{|j|\le J+3k}\,\frac{|c_{k,j}(t)|}{\sqrt{|j|!}}\,e^{\beta_d|j|/2}\\[4pt]
&&\quad\qquad\qquad\qquad\times\qquad
\sum_{|p|=l+2-k}\,\hbar^{|p|/2}\ \|A(t)\|^{|p|}\
2^{|p|/2}\ \sqrt{\Gamma(|j|+\frac d2+|p|)}\,.\label{berlin3}
\eea
The number of terms that occur in the final sum of this expression
is $\ds \left(\begin{array}{c}l-k+d+1\\d-1\end{array}\right)$, and the terms
in that sum are increasing. Thus, (\ref{berlin3}) is bounded by
\bea\nonumber
&&M_1\,{C''''}^{1/2}\,\E^{-b^2/(12\|A(t)\|^2\hbar)}\,\delta^{-l-2}\,
\sum_{k=0}^{l-1}\,\hbar^{k/2}\,\delta^k\,
\sum_{|j|\le J+3k}\,\frac{|c_{k,j}(t)|}{\sqrt{|j|!}}\ e^{\beta_d|j|/2}
\\ \nonumber
&&\quad\times\qquad
\left(\begin{array}{c}l-k+d+1\\d-1\end{array}\right)\left( 2\,\hbar\,\|A(t)\|^2\right)^{(l+2-k)/2}\,
\sqrt{\,\Gamma(|j|+\frac d2+l+2-k)}\\ \nonumber
&=&M_1\,{C''''}^{1/2}\,\E^{-b^2/(12\|A(t)\|^2\hbar)}\,\delta^{-l-2}\,
\left( 2\,\hbar\,\|A(t)\|^2\right)^{\frac l2+1}\
\sum_{k=0}^{l-1}\,\delta^k\,\left( 2\,\|A(t)\|^2\right)^{-k/2}\,\\ \nonumber
&&\quad\times\qquad
\left(\begin{array}{c}l-k+d+1\\d-1\end{array}\right)\sum_{|j|\le
  J+3k}\,
\frac{|c_{k,j}(t)|}{\sqrt{|j|!}}\ e^{\beta_d|j|/2}\
\sqrt{\,\Gamma(|j|+\frac d2+l+2-k)}\,.
\eea
Applying the Schwartz inequality to the sum over $j$, we see that this
expression is bounded by
\bea\nonumber
&&M_1\,{C''''}^{1/2}\,\E^{-b^2/(12\|A(t)\|^2\hbar)}\,\delta^{-l-2}\,
\left( 2\,\hbar\,\|A(t)\|^2\right)^{\frac l2+1}\
\sum_{k=0}^{l-1}\,\delta^k\,\left( 2\,\|A(t)\|^2\right)^{-k/2}\,
\\ \nonumber
&&\quad\qquad\times\qquad
\left(\begin{array}{c}l-k+d+1\\d-1\end{array}\right)\,\|c_k(t)\|
\left(\sum_{|j|\le J+3k}\,\frac{e^{\beta_d|j|}}{|j|!}\,
\Gamma(|j|+\frac d2+l+2-k)\,\right)^{1/2}.
\eea
The number of terms that occur in the final sum of this expression
is $\ds \left(\begin{array}{c}J+3k+d\\d\end{array}\right)$, and the terms
in that sum are increasing. Thus, the expression is bounded by
\bea\nonumber
&&M_1\,{C''''}^{1/2}\,\E^{-b^2/(12\|A(t)\|^2\hbar)}\,\delta^{-l-2}\,
\left( 2\,\hbar\,\|A(t)\|^2\right)^{\frac l2+1}\
\sum_{k=0}^{l-1}\,\delta^k\,\left( 2\,\|A(t)\|^2\right)^{-k/2}\,
\\ \nonumber
&&\ \times
\left(\begin{array}{c}l-k+d+1\\d-1\end{array}\right)\|c_k(t)\|
\left[\left(\begin{array}{c}J+3k+d\\d\end{array}\right)
\frac{e^{\beta_d(J+3k)}}{(J+3k)!}\Gamma(J+2k+\frac d2+l+2)\right]^{1/2}.
\eea
We now apply the estimate of $\|c_k(t)\|$ from Corollary \ref{cor} to
bound this by
\bea\nonumber
&&M_1\,{C''''}^{1/2}\,\E^{-b^2/(12\|A(t)\|^2\hbar)}\delta^{-l-2}\,
\left( 2\,\hbar\,\|A(t)\|^2\right)^{\frac l2+1}\\[4pt] \nonumber
&&\qquad\times\quad\sum_{k=0}^{l-1}\,\delta^k\,
\left( 2\,\|A(t)\|^2\right)^{-k/2}\,
\left(\begin{array}{c}l-k+d+1\\d-1\end{array}\right)\,
\left(\begin{array}{c}J+3k+d\\d\end{array}\right)^{1/2}
\\[4pt] \nonumber
&&\qquad\qquad\times\quad
e^{\beta_d(J+3k)/2}\,\frac{D_2^kD_3^k(1+D_1D_2^2|t|)^k}{k!}
\left(\frac{\Gamma(J+2k+\frac d2+l+2)}{J!}\right)^{1/2}\\[7pt] \nonumber
&\le&M_1{C''''}^{1/2}\E^{-b^2/(12\|A(t)\|^2\hbar)}\delta^{-l-2}
\left( 2\,\hbar\,\|A(t)\|^2\right)^{\frac l2+1}
\frac{e^{\beta_dJ/2}}{\sqrt{J!}}\\[4pt] \nonumber
&&\qquad\times\quad\sum_{k=0}^{l-1}\,\delta^k\left( 2\,\|A(t)\|^2\right)^{-k/2}
\left(\begin{array}{c}l-k+d+1\\d-1\end{array}\right)\,
\left(\begin{array}{c}J+3k+d\\d\end{array}\right)^{1/2}
\\[4pt] \nonumber
&&\qquad\qquad\qquad\times\qquad
e^{3\beta_dk/2}\ \frac{D_2^kD_3^k(1+D_1D_2^2|t|)^k}{k!}\
\sqrt{\Gamma(J+2k+\frac d2+l+2)}.\\&& \label{berlin4}
\eea
We now employ the following inequalities that hold for some numbers $D_5$ and
$D_6$:
\bea\nonumber
\left(\begin{array}{c}l-k+d+1\\d-1\end{array}\right)
&\le&\left(\begin{array}{c}l+d+1\\d-1\end{array}\right),\\[4pt] \nonumber
\left(\begin{array}{c}J+3k+d\\d\end{array}\right)
&\le&\left(\begin{array}{c}J+3l-3+d\\d\end{array}\right),\\[4pt] \nonumber
e^{3\beta_dk/4}&\le&e^{3\beta_d(l-1)/4},\\[4pt] \nonumber
D_2^k\,D_3^k\,(1+D_1D_2^2|t|)^k&\le&D_5^{l-1},\\[7pt] \nonumber
\frac 1{k!}\ \sqrt{\Gamma(J+2k+\frac d2+l+2)}&\le&
\frac 1{(l-1)!}\ \sqrt{\Gamma(J+3l+\frac d2)},
\qquad\qquad\mbox{and}\\[4pt] \nonumber
\sum_{k=0}^{l-1}\,\delta^k\,\left(2\,\|A(t)\|^2\right)^{-k/2}
&\le&D_6^l.
\eea
We then see that (\ref{berlin4}) is bounded by
\bea\nonumber
&&M_1{C''''}^{1/2}\E^{-b^2/(12\|A(t)\|^2\hbar)}\,\delta^{-l-2}\,
\left(2\,\hbar\,\|A(t)\|^2\right)^{\frac l2+1}\,
\frac{e^{\beta_dJ/2}}{\sqrt{J!}}\,
\left(\begin{array}{c}l+d+1\\d-1\end{array}\right)\,
\\[5pt] \nonumber
&&\qquad\times\qquad
\left(\begin{array}{c}J+3l-3+d\\d\end{array}\right)^{1/2}
\E^{3\beta_d(l-1)/4}\,D_5^{l-1}\,D_6^l\,
\frac 1{(l-1)!}\,\sqrt{\Gamma(J+3l+\frac d2)}.\\ \label{berlin5}
\eea
We bound this expression by using the two inequalities
\bea\nonumber
\left(\begin{array}{c}l+d+1\\d-1\end{array}\right)&\le&
(l+d+1)^{d-1}\qquad\qquad\mbox{and}\\[4pt] \nonumber
\left(\begin{array}{c}J+3l-3+d\\d\end{array}\right)&\le&
(J+3l-3+d)^d.
\eea
Note that the right hand sides of these inequalities grow polynomially
with $l$.

Since $d$ is fixed, we conclude that (\ref{berlin5}), and hence,
(\ref{awful}) are bounded by a constant times
\be\label{explicitJ2}
\E^{-b^2/(12\|A(t)\|^2\hbar)}\
\frac{\E^{\gamma' J}}{\sqrt{J!}}\,\hbar^{l/2}\ \frac{\E^{\gamma l}}{(l-1)!}\
\sqrt{\Gamma(J+3l+\frac d2)},
\ee
for some positive $\gamma$, and $\gamma'$.
With $J$ fixed, we apply Stirling's formula to the factorial and
$\Gamma$ function to bound this
by another constant times
$$\E^{-b^2/(12\|A(t)\|^2\hbar)}\ \hbar^{l/2}\ e^{\gamma l}\
\frac{(3l)^{3l/2}}{l^l}\ =\
\left(e^{\gamma}\,3^{3/2}\right)^l\,(\hbar\,l)^{l/2+1}.$$
By choosing $l\ =\ \grintl\,g/\hbar\,\grintr$ for some sufficiently
small $g>0$,
this is bounded by a constant times $\ds \,e^{-\gamma_2/\hbar}$.

This implies the lemma. \ep

\vskip .5cm
Theorem \ref{thm1} follows immediately from Lemmas (\ref{inside}) and
(\ref{outside}) with $G\,=\,\min\,\{\,G_1,\,G_2\,\}$.

\vskip .5cm
\section{Localization Estimates for the Wave Packets}\label{locsec}
\setcounter{equation}{0}
\setcounter{theorem}{0}

\vskip .25cm
In this section we show that our wave packets are localized near the classical
path. Given any $\epsilon>0$, we can choose the truncation parameter
$g>0$, such that our exponentially accurate wave packet is concentrated within
$\{\,x\,:\,|x-a(t)|<b \,\}$ up to an exponentially small error.\\

\noindent
{\bf Proof of Theorem \ref{thm2}:}
Let $\chi(x,t)$ be the characteristic function of the set
$\{\,x\,:\,|x-a(t)|>b\,\}$, and let $\psi(x,t,\hbar)$ be the result of
our construction with the series truncated with
$l(\hbar)\,=\,\grintl\,g/\hbar\,\grintr$.

We must prove
\be\label{lastfact}
\|\,\chi(\cdot,t)\,\psi(\cdot,t,\hbar)\,\|\ \le\ \exp\,\{\,-\Gamma/\hbar\,\},
\ee
for some $\Gamma>0$ when $g>0$ is sufficiently small.

The left hand side of (\ref{lastfact}) is bounded by
\be\label{rocky}
\sum_{k=0}^{l-1}\ \,\sum_{|j|\le J+3k}\ |c_{k,j}(t)|\ \|\,\chi(\cdot,t)\,
\phi_j(A(t),B(t),\hbar,a(t),\eta(t),\cdot)\,\|.
\ee
The norm in this sum has the form (\ref{target}), with $n=0$ and
$\tau=0$. Mimicking the estimation of (\ref{target}), we obtain the
estimate that corresponds to (\ref{berlin3}). We conclude that if $g>0$ is
sufficiently small, then
$$\|\,\chi(\cdot,t)\,\phi_j(A(t),B(t),\hbar,a(t),\eta(t),\cdot)\,\|\ \le\
\E^{-b^2/(12\|A(t)\|^2\hbar)}\ \E^{|j|\tilde{\beta}_d}$$
whenever $|j|\le J+3l(\hbar)-3$, for some $\tilde{\beta}_d$.

We use this and the Schwarz inequality to obtain, for some
constants $C_0$, $C_1$, $C_2$ and $C_3$
\bea\nonumber
&&\sum_{k=0}^{l-1}\ \sum_{|j|\le J+3k}\,\hbar^{k/2}\ |c_{k,j}(t)|\
\|\,\chi(\cdot,t)\,
\phi_j(A(t),B(t),\hbar,a(t),\eta(t),\cdot)\,\|\\[5pt]\nonumber
&\le&\sum_{k=0}^{l-1}\,\hbar^{k/2}\ \|\,c_k(t)\,\|\,
\E^{-b^2/(12\|A(t)\|^2\hbar)}\
\left(\,\sum_{|j|\le J+3k}\,\E^{|j|\tilde{2\beta}_d}\,\right)^{1/2}
\\[5pt]
&\le&\E^{-b^2/(12\|A(t)\|^2\hbar)}D_1D_2^2t\sum_{k=0}^{l-1} \hbar^{k/2}
\left(\frac{(J+3k)!}{J!}\right)^{1/2}\frac{C_0^k D_5^k}{k!}\,
\E^{\tilde{\beta}_d J}(J+3k+d)^{d/2}   \label{explicitJ3}
\\[5pt]
&\le&\E^{-b^2/(12\|A(t)\|^2\hbar)}\,C_1\,\sum_{k=0}^{\infty}\,
\left(\hbar C_2 k \right)^{k/2}\ \leq\ C_3\,\E^{-b^2/(12\|A(t)\|^2\hbar)},
\label{willy}
\eea
provided $g$ is small enough that
$\ds \hbar\,C_2\,k\ \leq\ C_2\,g\ <\ 1$ is satisfied.\qquad \ep

\vskip .5cm
\section{Ehrenfest Time Scale}\label{longtime}
\setcounter{equation}{0}
\setcounter{theorem}{0}

\vskip .25cm
In this section we consider the accuracy of our construction
when we allow $T$ to grow as $\hbar\ra 0$. Since the results stated
in Theorem \ref{thm4} and the method of proof
are basically equivalent to those of \cite{hagjoy3}, we will be
rather sketchy. \\

\noindent
{\bf Proof of Theorem \ref{thm3}}:\quad The first point to notice is
that the since potential is bounded from below, energy conservation
implies that $a(t)$ grows at most linearly with time. The exponential
bound on the potential then implies the existence of
$\widetilde{D}_1>0$ and $v>0$ such that the quantity (\ref{D1def}) is
bounded by
$$
D_1(T)\ \leq\ \widetilde{D}_1\,\E^{v\tau T}.
$$
Similarly, the existence of the Lyapunov exponent $\lambda$
implies the existence of $\widetilde{D}_2$ such that the quantity
(\ref{D2def}) satisfies
$$
D_2(T)\ \leq\ \tilde{D}_1\,\E^{\lambda T}.
$$
It then remains for us to keep track of the time dependence
in the proof of Theorem \ref{thm1}.
In particular, the quantities (\ref{D4def}) and (\ref{D5def})
fulfill the following estimates, modulo a possible increase of $v$:
\bea\nonumber
D_4(T)&\leq&D_1(T),\\[4pt]\nonumber
D_5(T)&\leq&\widetilde{D}_5\,\E^{(v\tau + 2\lambda) T}.
\eea
Using these bounds, we get the existence of constants
$C$ and $D$, independent of time, such that (\ref{insideest1})
can be replaced by
\be\label{tdest1}
  \hbar^{-1}\int_0^T\,\|\chi_1(\cdot ,t)\xi_l(\cdot, t, \hbar )\|\,dt\ \leq\
  \hbar^{-1}\,D\,\E^{(2v\tau+3\lambda)T}\,\left(\,C\,l\,\hbar\,
  \E^{(6\lambda+2v\tau)T}\,\right)^{l/2}.
\ee
Thus, if we choose $l=g(T)/\hbar$, then (\ref{tdest1}) is bounded by
$$
  \hbar^{-1}\,D\,\E^{(2v\tau+3\lambda)T}\,\left(\,C\,g(T)\,
  \E^{(6\lambda+2v\tau)T}\,\right)^{g(T)/(2\hbar)},
$$
so that we need
$$
  g(T)\,\E^{(6\lambda+2v\tau)T}\ra 0 \mbox{ and } g(T)/\hbar\ra \infty.
$$
These demands are satisfied by the choices
\be\label{cgt}
 g(T)\,=\,\E^{-\kappa T}\qquad\mbox{and}\qquad T\,=\,T'\ln(1/\hbar),
\ee
provided
\be\label{ccgt}
6\lambda+2v\tau\,<\,\kappa\,<\,1/T'.
\ee
Note that the prefactor in (\ref{tdest1}) will be of order
$\hbar^{-\nu_1}$, for some finite $\nu_1$. It will thus play no role
since it follows from these considerations that there exists
$\gamma_1>0$, such that
$$
  \hbar^{-1}\int_0^T\,\|\chi_1(\cdot ,t)\xi_l(\cdot, t, \hbar )\|\,dt
\ =\ O(\hbar^{-\nu_1}\,\E^{-\gamma_1/\hbar^{1-\kappa T'}}).
$$
By a similar argument, we obtain the estimate corresponding
to (\ref{outsideest}) with other time--independent constants
$D'$ and $C'$:
\be\label{tdest2}
  \hbar^{-1}\int_0^T\,\|\chi_2(\cdot ,t)\xi_l(\cdot, t, \hbar )\|\,dt\ \leq\
  \hbar^{-1}\,\E^{-b^2/(12 \hbar \e^{2\lambda T})}\,D'\,\left(C'\,l\,\hbar\,
  \E^{(8\lambda+2v\tau)T}\right)^{l/2}.
\ee
Inserting our choices (\ref{cgt}) and constraints (\ref{ccgt}) in
 (\ref{tdest2}), it is elementary to see that
 there exists positive $\nu_2$ and $\gamma_2$, such that
$$
  \hbar^{-1}\int_0^T\,\|\chi_2(\cdot ,t)\xi_l(\cdot, t, \hbar )\|\,dt
\ =\ O(\hbar^{-\nu_2}\,\E^{-\gamma_2/\hbar^{1-2\lambda T'}}),
$$
which proves the Theorem. \ep

\vskip .5cm \noindent
{\bf Proof of Theorem \ref{thm4}}:\quad
Considerations similar to those in the second part of the
proof of Theorem \ref{thm3} show that
there exists constants $C_0$, $C_1$ independent of $T$ such that
\bea\nonumber
&&\|\chi_2(\cdot,t)\psi(\cdot,t,\hbar)\|\\ &&\leq\
\sum_{k=0}^{l-1}\,\hbar^{k/2}\,\|c_k(t)\|\,\left(\,\sum_{|j|\leq J+3k}\,
\|\chi_2(\cdot, t)\,\phi_j(A(t), B(t), \hbar, a(t), \eta(t),\cdot)\|^2\,
\right)^{1/2}\nonumber\\ \nonumber
&&\leq\ \E^{-b^2/(12\hbar\e^{2\lambda T})}\,\sum_{k=0}^{g(T)/\hbar}\,
\left(\hbar k \E^{(2v\tau +6\lambda )T } C_0 \right)^{k/2}\
\leq\ \E^{-b^2/(12 \hbar \e^{2\lambda T})}\,\sum_{k=0}^{\infty}\,
C_1^k,
\eea
where, by virtue of (\ref{cgt}) and (\ref{ccgt}), we can take
$C_1<1$. So, the Theorem holds with exponential decay of order
$\E^{-b^2/(12\hbar^{(1-2\lambda \kappa T')})}$. \ep

\vskip .5cm
\section{Scattering Theory}\label{scatt}
\setcounter{equation}{0}
\setcounter{theorem}{0}

\vskip .25cm
In this section we show our approximations are valid up to exponentially small
corrections in a scattering framework, provided the potential satisfies
hypothesis D.

\vskip .5cm \noindent
{\bf Proof of Theorem \ref{thm5}}:\quad
First note that equations (\ref{clasflow}) together with
$$
  \E^{-itH_0(\hbar)/\hbar}\,\phi_j(A, B, \hbar, a, \eta, x)\ =\
\E^{it\eta^2/(2\hbar)}\,\phi_j(A+tiB, B, \hbar, a+t\eta, \eta, x)
$$
for any $j\in\N^d$ imply that as $t\ra\pm\infty$,
$$
\E^{itH_0(\hbar)/\hbar}\ \E^{iS(t)/\hbar}\,
\phi_j(A(t), B(t), \hbar, a(t), \eta(t), x)\,\ra\,
\E^{iS_{\pm}/\hbar}\,\phi_j(A_{\pm}, B_{\pm}, \hbar, a_{\pm}, \eta_{\pm}, x)
$$
with $S_-=0$, for any $j\in\N^d$.
Moreover, using (\ref{clasflow}) and the property
$$\min(|v|,1)\,\langle t\rangle\ \leq\ \langle t v\rangle\ \leq\
\max(|v|,1)\,\langle t\rangle,
$$
for any $v\in\R^d$ and any $t\in\R$, with
$\langle t\rangle\,=\,\sqrt{1+t^2}$, we get
the existence of $\tilde{c}_0>0$ and $\tilde{c}_1>0$ depending on the
asymptotic data $(a_{\pm}, \eta_{\pm})$, such that
\be\label{scatpot}
 \left|\,\frac{D^m V(a(t))}{m!}\,\right|\ \leq\
\frac{\tilde{c}_0\,\tilde{c}_1^{|m|}}{\langle t\rangle^{\beta+|m|}}
\ee
for large times. This estimate together with (\ref{linflow}) and Lemma
\ref{opnorm} yields the following estimate on the operator
$K_k(t)\,P_{|j|\leq n}$ defined in (\ref{Kkdef}):
\bea\nonumber
\| K_k(t)\,P_{|j|\leq n}\|
&=&\left\|\,
\sum_{|m|=k}\,\frac{D^m V(a(t))}{m!}\,X(t)^m\,P_{|j|\leq n}\,\right\|\\[4pt]
&\leq&
\pmatrix{d-1+k \cr d-1}\ \sqrt{\frac{(n+k)!}{n!}}\
\frac{\tilde{c}_0\,\tilde{c}_2^{k}}{\langle t\rangle^{\beta}}\,,\label{asesk}
\eea
where $\tilde{c}_2$ depends on the asymptotic data
$(a_{\pm}, \eta_{\pm}, A_{\pm}, B_{\pm})$ and the binomial coefficient gives the number
of multi-indices of order $k$. At the possible cost of an increase in the
constants, we may assume this estimate is valid for all $t\in\R$.

This estimate shows in particular that $K_k(t)\,P_{|j|\leq n}$ is
integrable in time.
From this, it is easy to check inductively that the
solutions $c_n(t)$ to the equations (\ref{ordern}) have limits
as $|t|\ra \infty$.

The asymptotic values of the coefficients $c_n(t)$ at infinity allow us
to define the asymptotic states $\Phi_{\pm}(A_{\pm}, B_{\pm}, \hbar, a_{\pm},
\eta_{\pm}, x )$ by (\ref{defastat}) with initial conditions at
$-\infty$ characterized by arbitrary normalized coefficients that satisfy
\bea\nonumber
c_{0,j}(-\infty)&=&0,\quad\mbox{for }|j|>J,\quad\mbox{and}\\ \label{asincon}
c_{n,j}&=&0,\quad\mbox{for } n=1,2,\cdots,\mbox{ and all } j\in\N^d.
\eea
Thus, our approximate solution
$$
\psi(x,t,\hbar)\ =\ e^{iS(t)/\hbar}
\sum_{|j|\le J+3g/\hbar-3}\,c_j(t,\hbar)\,\phi_j(A(t),B(t),\hbar,a(t),\eta(t),x)
$$
has the asymptotic property as $t\ra\pm\infty$,
\be\label{aslimop}
  \E^{itH_0(\hbar)/\hbar}\,\psi(x,t,\hbar)\,\ra\,\Phi_{\pm}(A_{\pm},
  B_{\pm}, \hbar, a_{\pm}, \eta_{\pm}, x ).
\ee
We prove below that
\bea
&&\|\psi(x,t,\hbar)\,-\,\lim_{s\ra\-\infty}\,\E^{i(t-s)H(\hbar)/\hbar}\,
  \E^{isH_0(\hbar)/\hbar}\,\Phi_{-}(A_{-},
  B_{-}, \hbar, a_{-}, \eta_{-}, x )\|\nonumber\\[3pt]
  &=&\|\psi(x,t,\hbar)\,-\,\E^{itH(\hbar)/\hbar}\,\Omega^+(\hbar)\,
  \Phi_{-}(A_{-},B_{-}, \hbar, a_{-}, \eta_{-}, x )\|\nonumber\\[5pt]
  &=&O(\E^{-\gamma/\hbar}),\nonumber
\eea
uniformly for $t\in\R$. Thus, making use of (\ref{aslimop}), we have
\bea
&&\lim_{t\ra +\infty}\,\|\E^{itH_0(\hbar)/\hbar}\,\psi(x,t,\hbar)\,-\,
\E^{itH_0(\hbar)/\hbar}\,\E^{itH(\hbar)/\hbar}\,\Omega^+(\hbar)
\,\Phi_{-}(A_{-},B_{-},\hbar, a_{-}, \eta_{-}, x )\|\nonumber\\[3pt]
&=&\|\Phi_{+}(A_{+},B_{+},\hbar, a_{+}, \eta_{+}, x )\,-\,
\Omega^-(\hbar)^*\,\Omega^+(\hbar)\,
\Phi_{-}(A_{-},B_{-},\hbar, a_{-}, \eta_{-}, x )\|\nonumber\\[3pt]
&=&\|\Phi_{+}(A_{+},B_{+},\hbar, a_{+}, \eta_{+}, x )\,-\,S(\hbar)\,
\Phi_{-}(A_{-},B_{-}, \hbar, a_{-}, \eta_{-}, x )\|\nonumber\\[5pt]
&=&O(\E^{-\gamma/\hbar}).\nonumber
\eea

Hence, we need only show that the estimate on $\xi_l(x,t,\hbar)$
corresponding to our approximation yields an exponentially small
correction term after choosing $l=g/\hbar$ for sufficiently small $g$,
uniformly for $t\in\R$.

We mimic Section 5 to get estimates on the coefficients
\be\label{excoscat}
c_k(t)\ =\ \sum_{p=1}^k\ c_k^{[p]}(t)
\ee
starting with
$$
c_{0,j}(t)=c_{0,j}(-\infty),\qquad c_{0,j}(-\infty)=0\quad
\mbox{if }|j|>J,\qquad
\mbox{and}\qquad\|c_0(-\infty)\|\,=\,1.
$$
We note that the number of components of the vectors $c_k^{[p]}(t)$
is the same as in (\ref{yeah2}) and that the combinatorics associated with
the $n$ and $p$ dependence of the estimates is identical to that performed
in Section 5.

Hence, with $D_3=\pmatrix{d+2\cr d-1}$, at first order we have
$$
\|c_1(t)\|\ =\ \|c_1^{[1]}(t)\|\ \leq\
D_3\ \sqrt{\frac{(J+3)!}{J!}}\ \tilde{c}_0\,
\tilde{c}_2^3\ \int_{-\infty}^t\,\langle s\rangle^{-\beta}\,ds.
$$
At second order, we obtain $c_2(t)=c_2^{[1]}(t)+c_2^{[2]}(t)$, where
\bea\nonumber
  \|c_2^{[1]}(t)\|&\leq&D_3^2\ \sqrt{\frac{(J+4)!}{J!}}\ \tilde{c}_0\,
\tilde{c}_2^4\,\int_{-\infty}^t\,\langle s\rangle^{-\beta}\,ds
\qquad\mbox{and}\\[5pt]
\|c_2^{[2]}(t)\|&\leq&D_3^2\ \sqrt{\frac{(J+6)!}{J!}}\ \tilde{c}_0^2\,
\tilde{c}_2^6\,\int_{-\infty}^t\,ds_1\,\langle s_1\rangle^{-\beta}\,
\int_{-\infty}^{s_1}\,ds_2\,\langle s_2\rangle^{-\beta}.\nonumber
\eea
At third order, we obtain
$c_3(t)=c_3^{[1]}(t)+c_3^{[2]}(t)+c_3^{[3]}(t)$, where
\bea\nonumber
\|c_3^{[1]}(t)\|&\leq&D_3^3\ \sqrt{\frac{(J+5)!}{J!}}\ \tilde{c}_0\,
\tilde{c}_2^5\,\int_{-\infty}^t\,\langle s\rangle^{-\beta}\,ds,\\[5pt]
\|c_3^{[2]}(t)\|&\leq&2\,D_3^3\ \sqrt{\frac{(J+7)!}{J!}}\ \tilde{c}_0^2\,
\tilde{c}_2^7\,\int_{-\infty}^t\,ds_1\,\langle s_1\rangle^{-\beta}\,
\int_{-\infty}^{s_1}\,ds_2\,\langle s_2\rangle^{-\beta},
\qquad\mbox{and}\nonumber\\
\|c_3^{[3]}(t)\|&\leq&D_3^3\ \sqrt{\frac{(J+9)!}{J!}}\ \tilde{c}_0^3\,
\tilde{c}_2^9\int_{-\infty}^t ds_1\,\langle s_1\rangle^{-\beta}
\int_{-\infty}^{s_1} ds_2\,\langle s_2\rangle^{-\beta}\int_{-\infty}^{s_2}
ds_3\,\langle s_3\rangle^{-\beta}.\nonumber
\eea
Using the identity
$$
\int_{-\infty}^t\,ds_1\,\langle s_1\rangle^{-\beta}
\int_{-\infty}^{s_1}\,ds_2\,\langle s_2\rangle^{-\beta}\,\cdots\,
\int_{-\infty}^{s_{n-1}}\,ds_n\,\langle s_n\rangle^{-\beta}
\ =\ \frac 1{n!}\
\left(\,\int_{-\infty}^t\,\langle s\rangle^{-\beta}\,ds\,\right)^n,\nonumber
$$
we get estimates identical to (\ref{c11est}), (\ref{c21est}), (\ref{c22est}),
(\ref{c31est}), (\ref{c32est}), (\ref{c33est}).

It is easy to check that the induction can be carried out exactly as in
Section 5 to give an analog of Corollary 5.3 that states

\begin{lemma}
Assume the decay hypothesis D. The expansion coefficients
(\ref{excoscat}) satisfying (\ref{asincon}) obey the following estimates:
\bea\nonumber
c_{k,j}(t)&=&0,\quad\mbox{ whenever }\quad |j|>J+3k,\\[4pt]\nonumber
\|c_k^{[p]}(t)\|&\leq&\pmatrix{k-1\cr p-1}\ D_3^k\
\sqrt{\frac{(J+3k)!}{J!}}\ \tilde{c}_0^p\ \tilde{c}_2^{k+2p}\
\frac{\left(\int_{-\infty}^t ds\,\langle s\rangle^{-\beta}\right)^p}{k!}
\eea
for $p\leq k$, and
$$
\|c_k(t)\|\,\leq\,\sqrt{\frac{(J+3k)!}{J!}}\ \frac{\tilde{c}_2^k\,D_3^k}{k!}\,
\left(1+\tilde{c}_0\tilde{c}_2^2
\int_{-\infty}^t ds\,\langle s\rangle^{-\beta}\right)^{k-1}\
\tilde{c}_0\ \tilde{c}_2^2\ \int_{-\infty}^t ds\,\langle s\rangle^{-\beta}.
$$
\end{lemma}

\vskip .3cm
Our next task is to estimate the norm of $\xi_l(x,t)$. We again consider
separately the errors near the classical orbit and those far from the orbit.
Let $b(t)$ be a real valued function that satisfies
\be\label{defbt}
\frac{\langle a(t)\rangle}{4}\ \leq\ b(t)\ \leq\ \frac{\langle a(t)\rangle}{2},
\ee
for all $t\in \R$.
We define $\chi_1(x,t)$ to be the characteristic function of
$\{x\, :\, |x-a(t)|\leq b(t)\}$ and $\chi_2(x,t)=1-\chi_1(x,t)$.
Then, for some constants $\tilde{c}_3$ and $\tilde{c}_4$ and
any $t\in\R$, we have
\bea
\left|\,\frac{D^m V(\zeta_m(x,a(t)))}{m!}\,\chi_1(x,t))\,\right| &\leq&
\frac{v_0\ v_1^{|m|}}{\langle\zeta_m(x,a(t)))\rangle^{\beta+|m|}}\ \chi_1(x,t)
\nonumber\\[4pt]
&\leq&\frac{\tilde{c}_3\ \tilde{c}_4^{|m|}}{\langle t\rangle^{\beta+|m|}},
\nonumber
\eea
since for large times on the support of $\chi_1$,
$$
|\zeta_m(x,a(t)))|\ \geq\ |a(t)|/4.
$$
Therefore, for some constants $\tilde{c}_5$ and $\tilde{c}_6$,
$$
\left\|\,\chi_1(x,t)\,\frac{D^m V(\zeta_m(x,a(t)))}{m!}\,
X(t)^m\,P_{|j|\leq n}\,\right\|\
\leq\ \sqrt{\frac{(n+|m|)!}{n!}}\
\frac{\tilde{c}_5\,\tilde{c}_6^{|m|}}{\langle t\rangle^{\beta}},
$$
for any $t\in\R$.
We now mimic the manipulations performed in Section 6 to get
\bea\label{chi1}
&&\|\,\chi_1(x,t)\,\xi_l(x,t,\hbar)\,\|\\[6pt]\nonumber &\le&
\Bigg\|\,\sum_{k=0}^{l-1}\,\sum_{|m|=l+2-k}\,\hbar^{k/2}\,\,\chi_1(x,t)\
\frac{(D^mV)(\zeta_{m}(x,a(t))}{m!}\ \hbar^{|m|/2}\,(x-a(t))^m\,\hbar^{-|m|/2}\\
\nonumber &&\qquad\qquad\qquad\qquad\times\quad
P_{|j|\leq J+3k}\ \sum_{|j|\le J+3k}\,c_{k,j}(t)\,
\phi_j(A(t),B(t),\hbar,a(t),\eta(t),x)\,\Biggr\|\\[6pt]\nonumber
&\le&\sum_{k=0}^{l-1}\ \hbar^{(l+2)/2}\,\pmatrix{d-1+l+2-k\cr d-1}\ \|c_k(t)\|
\\ \nonumber
&&\qquad\qquad\times\qquad
 \max_{\{m\,:\ |m|=l+2-k\} }\ \left\|\,\chi_1(x,t)\,\frac{D^m
V(\zeta_m(x,a(t)))}{m!}\,X(t)^m\,P_{|j|\leq J+3k}\,\right\|\nonumber\\
&\le&\sum_{k=0}^{l-1}\,\hbar^{(l+2)/2}\,\pmatrix{d-1+l+2-k\cr d-1}
\sqrt{\frac{(J+2k+l+2)!}{J!}}\ \frac{\tilde{c}_5\,\tilde{c}_6^{l+2-k}}
{\langle t\rangle^{\beta}} \nonumber\\
&&\qquad\qquad\qquad\times\quad
 \frac{\tilde{c}_2^k\,D_3^k}{k!}
 \left(1+\tilde{c}_0\tilde{c}_2^2
 \int_{-\infty}^t ds\,\langle s\rangle^{-\beta}\right)^{k-1}\,
 \tilde{c}_0\,\tilde{c}_2^2\,
 \int_{-\infty}^t ds\,\langle s\rangle^{-\beta}. \nonumber
\eea
Making use of (\ref{Alptraum}), the definition of $D_3$, and introducing
another constant
$$
\tilde{c}_7\ =\ \left(\,1\,+\,\tilde{c}_0\,\tilde{c}_2^2\,I\,\right)
\tilde{c}_2/\tilde{c}_6,\qquad\mbox{ where }\qquad
I\ =\ \int_{-\infty}^{\infty}\,\langle s\rangle^{-\beta}\,ds,
$$
(\ref{chi1}) is bounded by
\bea\label{first}
&&\frac{\tilde{c}_5\,\tilde{c}_0\,(\tilde{c}_6\,\tilde{c}_2)^2\,I}
{\sqrt{J!}\,\langle t\rangle^{\beta}}\
\hbar^{(l+2)/2}\ (\tilde{c}_6\,D_3)^l\ \sum_{k=0}^{l-1}\ \tilde{c}_7^k\
\frac{\sqrt{(J+2k+l+2)!}}{k!} \nonumber\\
&\leq&
\frac{\tilde{c}_5\,\tilde{c}_0\,(\tilde{c}_6\,\tilde{c}_2)^2\,I}
{\sqrt{J!}\,\langle t\rangle^{\beta}}\
\hbar^{(l+2)/2}\ (\tilde{c}_6\,D_3)^l\ \sum_{k=0}^{l-1}\ \tilde{c}_7^k\
\frac{\sqrt{(J+3l)!}}{(l-1)!}\nonumber\\
&\leq&\frac{\tilde{c}_5\,\tilde{c}_0\,(\tilde{c}_6\,\tilde{c}_2)^2\,I}
{(\tilde{c}_7-1)\,\sqrt{J!}\,\langle t\rangle^{\beta}}\
\hbar^{(l+2)/2}\ (\tilde{c}_6\,D_3\tilde{c}_7)^l\
\frac{\sqrt{(J+3l)!}}{(l-1)!}\nonumber\\
&=&\frac{\tilde{c}_5\,\tilde{c}_0\,(\tilde{c}_6\,\tilde{c}_2)^2\,I}
{(\tilde{c}_7-1)\sqrt{J!}\,\langle t\rangle^{\beta}}\
\hbar^{(l+2)/2}\ \left(D_3
\left(1+\tilde{c}_0\tilde{c}_2^2 I \right)\tilde{c}_2\right)^l\
\frac{\sqrt{(J+3l)!}}{(l-1)!}.
\eea
This estimate is integrable for $t\in\R$ and yields a bound that implies
exponential decay in $\hbar$ by the optimal truncation technique.

\vskip .25cm
We now come to the estimate of $\|\chi_2(x,t)\xi_l(x,t)\|$, which is a little
bit more elaborate. The difficulty stems from a lack of sufficient information
on the position of $\zeta_m(x,a(t))$.
So, instead of the usual Taylor series error formula, we use the definition
$$
W^{(q)}_{a(t)}(x)=V(x)\,-\,\sum_{|m|\le q}\,\frac{D^mV(a(t))}{m!}\,(x-a(t))^m.
$$
We ultimately use $q=l+1-k$, where $k=0,1,\cdots,l-1$.
Our proof requires the space dimension to satisfy $d\geq 3$ in order to obtain
integrability in $t$.

Consider the following integral
\bea
N^2&=&\int_{\scriptsize{\R}^d}\,\chi_2(x,t)^2\,|\phi_j(A(t),B(t),\hbar, a(t),
\eta(t), x)|^2\,V(x)^2\,dx\nonumber\\[4pt]\nonumber
&=&\int_{|z|\geq b(t)}\,|\phi_j(A(t),B(t),\hbar, 0, 0, z)|^2\,V(z+a(t))^2\,dz.
\eea
We use the formula
$$
|\phi_j(A(t),B(t),\hbar, 0, 0, z)|\ =\ \frac{\E^{-(|A|^{-1}(t)z)^2/2\hbar}\,
|{\cal H}_j(A;|A(t)|^{-1}\hbar^{1/2}z)|}
{\sqrt{j!\,2^{|j|}\,\pi^{d/2}\,|\det A(t)|\,\hbar^{d/2}}},
$$
the asymptotic behavior (\ref{linflow}), and
the following estimate, which is valid on the support of $\chi_2$,
$$
(|A(t)|^{-1}z)^2\ \geq\ \frac{z^2}{\|A(t)\|^2}\ \geq\
\frac{b^2(t)}{\|A(t)\|^2}\ \geq\ \frac{\langle a(t)\rangle^2}{16\,\|A(t)\|^2}
$$
to obtain the bound
$$
N^2\ \leq\ \E^{-\tilde{b}/\hbar}\int_{|z|\geq b(t)}\,
\frac{\E^{-(|A(t)|^{-1}z)^2/2\hbar}\,
|{\cal H}_j(A;|A(t)|^{-1}\hbar^{1/2}z)|^2}
{j!\,2^{|j|}\,\pi^{d/2}\,|\det A(t)|\,\hbar^{d/2}}\ V(z+a(t))^2\,dz,
$$
for some finite, positive $\tilde{b}$.
Note that this estimate has a uniform exponentially decreasing prefactor.

As in Section \ref{truncsec}, we use spherical coordinates and the
decomposition (\ref{chgbasis})
$$
\Omega_j(y)\ =\ \sum_{\{q,n,m:\,2n+q=|j|\}} d_{j,q,n,m}\,\psi_{q,n,m}(r,\omega),
$$
where $\ds\sum_{\{q,n,m:\,2n+q=|j|\}} |d_{j,q,n,m}|^2\ =\ 1$
and $\ds\Omega_j(y)\ =\ \sqrt{\frac{1}{2^{|j|}\,j!\,\pi^{d/2}}}\
{\cal H}_j(A;\,y)\,\E^{-y^2/2}$.
This leads to the estimate
$$
N^2\ \leq\ \E^{-\tilde{b}/\hbar}\sum_{\{q,n,m:\,2n+q=|j|\}}
\,\int_{|z|\geq b(t)}\,
\frac{\E^{+(|A(t)|^{-1}z)^2/2\hbar}\,|\,\psi_{q,n,m}(r_z,\omega_z)|^2}
{|\det A(t)|\,\hbar^{d/2}}\ V(z+a(t))^2\,dz,
$$
where the spherical coordinates $(r_z,\omega_z)$ describe
the vector $\hbar^{1/2}\,|A(t)|^{-1}\,z$.

We choose $p>2$, such that $d/\beta\,<\,p\,<\,d$, and define $s>2$ by
$1/s\,+\,1/p\,=\,1/2$. Applying H\"older's inequality, we get the bound
$$
N^2\ \leq\ \frac{\E^{-\tilde{b}/\hbar}}{|\det A(t)|\,\hbar^{d/2}}
\sum_{\{q,n,m:\,2n+q=|j|\}}
\|V\|_p^2\left(\int_{|z|\geq b(t)}
\E^{+(|A(t)|^{-1}z)^2s/4\hbar}\,|\psi_{q,n,m}(r_z,\omega_z)|^s dz\right)^{2/s}.
$$
We need to bound the integral in this expression. We change variables to
$y=|A(t)|^{-1}z/\hbar^{1/2}$ and use
the estimate $\left|\,|A(t)|y\,\right|\,\leq\,b(t)/\sqrt{\hbar}$,
which is valid when $|y|\leq b(t)/(\|A(t)\|\sqrt{\hbar})$. This yields
\bea
&&\int_{|z|\geq b(t)}\
\E^{+(|A(t)|^{-1}z)^2s/4\hbar}\ |\psi_{q,n,m}(r_z,\omega_z)|^s\,dz
\nonumber \\[4pt]
&\leq&\int_{|y|\geq\frac{b(t)}{\|A(t)\|\sqrt{\hbar}}}\,
\E^{+y^2s/4}\ |\psi_{q,n,m}(r,\omega)|^s\ |\det A(t)|\,\hbar^{d/2}\,dy,
\label{integrale}\eea
where the spherical coordinates $(r,\omega)$ now describe the vector
$y$. Note that we have used $\det(|A|)=|\det(A)|$, which follows from
$A=U_A\,|A|$, where $U_A$ is unitary.

Since $b(t)/\|A(t)\|$ has a strictly positive infimum $\bar{b}$,
and we ultimately choose $l\simeq g/\sqrt{\hbar}$, with $g$ arbitrarily small,
we can assume the integration in (\ref{integrale}) is within
the classically forbidden region where Lemma 4.2 applies,
for all indices $\{q,n,m:\,2n+q=|j|\}$ of interest.

Hence, manipulations similar to those performed in
Section \ref{truncsec}, show that (\ref{integrale}) is bounded above by
\bea
&&\frac{|\det A(t)|\,\hbar^{d/2}\,2^{s/2}}{\Gamma(q+n+d/2)^{s/2}\,n!^{s/2}}\
\int_{S^{d-1}}\,d\omega
\,|Y_{q,m}|^s\,\int_{\bar{b}/\sqrt{\hbar}}^{\infty}\,dr\,r^{d-1+sq+2sn}
\,\E^{-sr^2/4}\nonumber\\[4pt]
&\leq&\frac{|\det A(t)|\,\hbar^{d/2}\,(2/s)^{d+s|j|}\,2^{s/2}}
{\Gamma(q+n+d/2)^{s/2}\,n!^{s/2}}\
\int_{S^{d-1}}\,d\omega\,|Y_{q,m}|^s\,\int_0^{\infty}\,dz\,z^{d-1+s|j|}
\,\E^{-z^2}\nonumber\\[4pt]\nonumber
&\leq&\frac{|\det A(t)|\,\hbar^{d/2}\,(2/s)^{d+s|j|}\,2^{s/2}}
{\Gamma(q+n+d/2)^{s/2}\,n!^{s/2}}\
\int_{S^{d-1}}\,d\omega\,|Y_{q,m}|^s\,\Gamma\left(\frac{d+s|j|}{2}\right)/2.
\eea
This implies the estimate
\bea
N^2&\leq&\frac{\E^{-\tilde{b}/\hbar}\,\|V\|_p^2\,(2/s)^{2(d/s+|j|)}
\,\Gamma\left((d+s|j|)/2\right)^{2/s}}
{2^{2/s-1}\,|\det A(t)|^{1-2/s}\,\hbar^{d/2-d/s}}\nonumber\\[4pt]\nonumber
&&\qquad\qquad\qquad\qquad\qquad\times\qquad\left(\sum_{\{q,n,m:\,2n+q=|j|\}}\
\frac{\int_{S^{d-1}}\,d\omega\,|Y_{q,m}|^s}
{\Gamma(q+n+d/2)^{s/2}\,n!^{s/2}}\right)^{2/s}.
\eea
We bound the integral in this expression by using the following crude
lemma. Its proof is at the end of this section.

\vskip .3cm
\begin{lemma}\label{stupid}
For some constants $M_0$ and $M_1$, we have
\be\label{spharm}
\left|\,Y_{q,m}(\omega)\,\right|\ \le\ M_0\ M_1^q.
\ee
\end{lemma}
\vskip .3cm
We use this and the inequalities $q\leq |j|$ and
$\pmatrix{|j|\cr n}\leq 2^{|j|}$ to estimate
\bea
&&\left(\,\sum_{\{q,n,m:\,2n+q=|j|\}}\,
\frac{\int_{S^{d-1}}\,d\omega\,|Y_{q,m}|^s}{\Gamma(q+n+d/2)^{s/2}\,n!^{s/2}}
\,\right)^{2/s}\nonumber\\[4pt]
&\leq&\frac{M_0^2\,|S^{d-1}|^{2/s}\,(2dM_1^2)^{|j|}\,m_{|j|}^{2/s}}
{\pi^{d/2}}\ \left(\,\sum_{n=0}^{|j|/2}\,
\frac{1}{\Gamma(|j|-n+d/2)^{s/2}\,n!^{s/2}}\,\right)^{2/s}\nonumber\\[4pt]
&\leq&\frac{M_0^2\,|S^{d-1}|^{2/s}\,(2dM_1^2)^{|j|}\,m_{|j|}^{2/s}}
{\pi^{d/2}\,C''\,|j|!}\ \left(\,\sum_{n=0}^{|j|/2}\,
\pmatrix{|j|\cr n}^{s/2}\,\right)^{2/s}\nonumber\\[4pt]\nonumber
&\leq&\frac{|j|^{2/s}\,M_0^2\,|S^{d-1}|^{2/s}\,(4dM_1^2)^{|j|}\,m_{|j|}^{2/s}}
{\pi^{d/2}\,C''\,|j|!}.
\eea
Hence, for some constants $N_0$ and $N_1$, that depend on
$d$ and $s$ only,
$$
N\ \leq\ \E^{-\tilde{b}/2\hbar}\
\frac{\|V\|_p\,N_0\,N_1^{|j|}}{|\det A(t)|^{1/p}\,\hbar^{d/p}}.
$$
By our choice of $p$,
$$\frac{1}{|\det A(t)|^{1/p}}\ \simeq\ 1/\langle t\rangle^{d/p}
$$
is integrable.

For $k\leq l\simeq g/{\hbar}$, with sufficiently small $g$,
this last estimate allows us to bound the corresponding term in
$\|\xi_l(x,t)\chi_2(x,t)\|$ as follows (where the
$N_i$, $i=0,1,2,3\dots$ are constants):
\bea
&&\sum_{k=0}^{l-1}\,\hbar^{k/2}\,\left\|\,\sum_{|j|\leq J+3k}\,c_{k,j}(t)
\,\chi_2(x,t)\,V(x)\,\phi_j(A(t), B(t),\hbar,a(t),\eta(t),x)\,\right\|
\nonumber\\
&\leq&\sum_{k=0}^{l-1}\,\hbar^{k/2}\,\|c_k(t)\|\,\left(\,\sum_{|j|\leq J+3k}\,
\|V(x)\,\phi_j(A(t), B(t),\hbar,a(t),\eta(t),x)\|\,\right)^{1/2}\nonumber\\
&\leq&\sum_{k=0}^{l-1}\,\hbar^{k/2}\,\sqrt{\frac{(J+3k)!}{J!}}\
\frac{N_3^k}{k!}\ \E^{-\tilde{b}/2\hbar}\
\frac{\|V\|_p\,N_0}{|\det A(t)|^{1/p}\,\hbar^{d/p}}\,\left(\sum_{|j|\leq J+3k}
N_1^{2|j|}\right)^{1/2} \label{Jexplicit4}\\
&\leq&\frac{\E^{-\tilde{b}/2\hbar}\,N_4}{|\det A(t)|^{1/p}\,\hbar^{d/p}}
\ \sum_{k=0}^{l-1}\,k^{k/2}\,\hbar^{k/2}\,N_5^{k}\nonumber\\
&\leq&\frac{\E^{-\tilde{b}/2\hbar}\,N_6}{|\det A(t)|^{1/p}\,\hbar^{d/p}}.
\label{middle}
\eea

It remains for us to control integrals of the form
\bea
F^2(p)&=&\int_{\R^d}\chi_2^2(x,t)|\phi_j(A(t), B(t), \hbar, a(t), \eta(t), x)|^2
|D^pV(a(t))(x-a(t))^p|^2/(p!)^2dx \nonumber\\[4pt]
&\leq&\frac{\tilde{c}_0^2\,\tilde{c}_1^{2|p|}}{\langle t\rangle^{2(\beta+|p|)}}
\ \int_{|x-a(t)|\geq b(t)}\,(x-a(t))^{2|p|}\,
|\phi_j(A(t),B(t),\hbar,a(t),\eta(t),x)|^2\,dx\nonumber\\[6pt]\nonumber
&\leq&\frac{\tilde{c}_0^2\,\tilde{c}_2^{2|p|}\,\E^{-\tilde{b}/(2\hbar)}\,
\hbar^{|p|}}{\langle t\rangle^{2\beta}}\
\int_{|y|\geq\bar{b}/\sqrt{\hbar}}\,y^{2|p|}|\,{\cal H}_j(A; y)|^2\,
\E^{y^2/2}/(2^{|j|}\,j!\,\pi^{d/2})\,dy,
\eea
where we used the same type of estimates as above.
We bound the last integral in this expression by
using spherical coordinates and
noting that the integration region lies within the classically forbidden region,
if $g$ is sufficiently small. The integral is thus bounded by
$$
\sum_{\{q,n,m\,:\,2n+q=|j|\}}\,\frac{2^{2n+q+|p|+d/2-1}}{n!\,\Gamma(q+n+d/2)}\,
\Gamma(2n+q+|p|+d/2)\ \leq\ f_0\,f_1^{|p|}\,(|j|+|p|)!/|j|!.
$$
So, for some other constants we have
$$
F(p)^2\ \leq\ \frac{f_2\,f_3^{|j|}\,f_4^{|p|}\,\hbar^{|p|}\,
\E^{-\tilde{b}/(2\hbar)}\,(|j|+|p|)!}{\langle t\rangle^{2\beta}\,|j|!}.
$$

The corresponding sum in $\|\chi_2\xi_l\|$ is bounded by
\bea\nonumber
&&\left\|\,\sum_{k=0}^{l-1}\,h^{k/2}\,\sum_{|j|\leq J+3k}\,c_{k,j}(t)
\right.\\
&&\qquad\quad\times\quad\left.
\sum_{|p|\leq l+1-k}\,\chi_2(x,t)\,\phi_j(A(t),B(t),\hbar,a(t),\eta(t),x)\,
\frac{D^pV(a(t))\,(x-a(t))^p}{p!}\,\right\|\nonumber\\
&\leq&\sum_{k=0}^{l-1}\,h^{k/2}\,\sum_{|p|\leq l+1-k}\,\|c_k(t)\|\,
\left(\,\sum_{|j|\leq J+3k}\,F^2(p)\,\right)^{1/2}\nonumber\\
&\leq&\sum_{k=0}^{l-1}h^{k/2}f_5^k \frac{\sqrt{(J+3k)!}}{\sqrt{J!}\,k!}
\sum_{|p|\leq l+1-k}\left( \sum_{|j|\leq J+3k}  \frac{f_2 f_3^{J+3k}
f_4^{|p|} \hbar^{|p|}\E^{-\tilde{b}/(2\hbar)}(J+3k+|p|)!}
{\langle t\rangle^{2\beta}\, (J+3k)!}\right)^{1/2}\nonumber\\
&\leq&\frac{\E^{-\tilde{b}/(4\hbar)}}{\langle t\rangle^{\beta}}\
\frac{f_2^{1/2}\,\E^{\bar{\beta}J}}{\sqrt{J!}}\
\sum_{k=0}^{l-1}\,\frac{h^{k/2}\,f_6^k}{k!}\sum_{|p|\leq l+1-k}
\hbar^{|p|/2}\,f_7^{|p|}\,\sqrt{(J+3k+|p|)!},\label{sumsum}
\eea
where $\bar{\beta}$ is independent of $J$.
The last sum in this expression is bounded by
\be\label{unesum}
\sum_{r=0}^{l+1-k}\,(f_8\hbar^{1/2})^r\,\sqrt{(J+3k+r)!}
\ =\ \sum_{s=3k}^{l+1+2k}\,(f_8\,\hbar^{1/2})^s\,\sqrt{(J+s)!}\,
(f_8\,\hbar^{1/2})^{-3k}.
\ee
Since $s\leq l+1+2k\leq 3l+1\simeq g/{\hbar}$ and $g$ is small, we have
$$
(f_8\,\hbar^{1/2})^s\,\sqrt{(J+s)!}\ \leq\ (f_9\,\hbar^{1/2}\,s^{1/2})^s\ \leq\
\alpha(g)^s,
$$
where $\alpha(g) =f_{10} \sqrt{g}$ is smaller than one.
Furthermore, the sum (\ref{unesum}) is bounded by
$$
\frac{\alpha(g)^{3k}}{1-\alpha(g)}\ (f_8\,\hbar^{1/2})^{-3k}.
$$
From this we deduce that (\ref{sumsum}) is dominated by
a constant times
\bea\nonumber
\frac{\E^{-\tilde{b}/(4\hbar)}}{\langle t\rangle^{\beta}}\
\sum_{k=0}^{l-1}\,\frac{h^{k/2}\,f_6^k}{k!}\,2\,\alpha(g)^{3k}\,
(f_8\,\hbar^{1/2})^{-3k}
&=&2\,\frac{\E^{-\tilde{b}/(4\hbar)}}{\langle t\rangle^{\beta}}\
\sum_{k=0}^{l-1}\,\frac{\tilde{\alpha}(g)^k}{\hbar^k\,k!}\nonumber\\[4pt]
&\leq& 2\,\frac{\E^{-\tilde{b}/(4\hbar)}}{\langle t\rangle^{\beta}}\
\E^{\tilde{\alpha}(g)/\hbar}\nonumber\\[4pt]
&\leq&\frac{\E^{-\tilde{b}/(8\hbar)}}{\langle t\rangle^{\beta}},\label{last}
\eea
provided $g$ is small enough, since $\tilde{\alpha}(g)\simeq g^{3/2}$ as
$g\ra 0$, which is exponentially small and integrable in time.

Finally, gathering estimates (\ref{first}), (\ref{middle}) and
(\ref{last}), we get the existence of positive $\gamma$, $H$, $G$, and $C$,
such that $g<G$,\ $l(\hbar)=g/{\hbar}$,\ and $\hbar<H$ imply
$$
\int_{-\infty}^{\infty}\,dt\,\|\xi_l(x,t)\|/\hbar\ \leq\ C\,\E^{-\gamma/\hbar}.
$$
\vskip .3cm \noindent
{\bf Proof of Lemma \ref{stupid}:}\quad
Let $f\in{\cal S}$ be the function that is given in spherical coordinates by
$$
f(x)\ =\
\sqrt{\frac{2}{\Gamma(q+\frac d2)}}\,\ r^q\ e^{-r^2/2}\ Y_{q,m}(\omega).
$$
For integers $q>0$, the maximum absolute value of this function is
\be\label{stupid1}
\sqrt{\frac{2}{\Gamma(q+\frac d2)}}\,\ q^{q/2}\ e^{-q/2}\
\max_{\omega}\ |\,Y_{q,m}(\omega)\,|.
\ee

The function $f$ is a normalized eigenfunction of $-\Delta+x^2$ with
eigenvalue $E=2q+d$.\\
Its norm in the Sobolev space ${\cal H}_s$ for $s>0$ satisfies
\be\label{stupid2}
\|f\|_{{\cal H}_s}\ \leq\ C_1(s)\
\left(\,\|f\|\,+\,\|(-\Delta)^{s/2}f\|\,\right)\
\leq\ C_1(s)\ \left(1\,+\,(2q+d)^{s/2}\right),
\ee
for some $C_1(s)$.

If $s>d/2$, then $(1+|k|^2)^{-s/2}$ is in $L^2(\R^d)$. So,
by H\"older's inequality,
$$
|\,f(x)\,|\ \le\
(2\pi)^{-d/2}\ \left\|\,\widehat{f}(k)\,(1+|k|^2)^{s/2}\,\right\|\
\left\|\,(1+|k|^2)^{-s/2}\,\right\|\ =\
C_2(s)\ \|f\|_{{\cal H}_s}.
$$
This and (\ref{stupid2}) imply that (\ref{stupid1}) is bounded by
$\ds C_3(s)\ \left(1\,+\,(2q+d)^{s/2}\right)$. Thus,
$$\max_{\omega}\ \left|\,Y_{q,m}(\omega)\,\right|
\ \le\ C_3(s)\
\sqrt{\frac{\Gamma(q+\frac d2)}2}\,\
\left(1\,+\,(2q+d)^{s/2}\right)\ q^{-q/2}\ e^{q/2}.
$$
The lemma follows from this by an application of Stirling's formula.\ep

\vskip .5cm
\section{More General Coherent States}\label{initcond}
\setcounter{equation}{0}
\setcounter{theorem}{0}

\vskip .25cm
In this section we extend all the previous theorems of the paper to
allow initial conditions that are certain infinite linear combinations of the
$\phi_j$.

\vskip .25cm\noindent
{\bf Proof of Theorem \ref{thm6}}:\quad
The strategy is quite simple. Let $\ffi\in {\cal C}$ have
expansion $\ffi\,=\,c_j\,\phi_j(\,\un,\,\un,\,\hbar,\,0,\,0,\,x)$, and let
$$
\psi_0(x,0,\hbar)\,=\,(\Lambda_h(a,\eta)\ffi)(x)
$$
be our initial condition.
By construction,
$$
\psi_0(x,0,\hbar)\,=\,
\sum_{j\in\N^d}\,c_j\,\phi_j(\,\un,\,\un,\,\hbar,\,a,\,\eta,\,x).
$$
For $J>0$, we define
$$
\psi_J(x,0,\hbar)=\sum_{|j|\leq J}\,c_j\,
\phi_j(\,\un,\,\un,\,\hbar,\,a,\,\eta,\,x),
$$
and denote the approximation that arises from this initial condition by
$$
\psi_{\tilde{J}(J,\hbar)}(x,t,\hbar)\,=\,
\sum_{|j|\leq \tilde{J}(J,\hbar)}\,c_j(t,\hbar)\,
\phi_j(A(t),B(t),\hbar,a(t),\eta(t),x).
$$
We then have
\bea\nonumber
&&\E^{-itH(\hbar)/\hbar}\,\psi_0(0,\hbar)\\[7pt]\label{strategy}
&=&\E^{-itH(\hbar)/\hbar}\,
(\psi_0(0,\hbar)-\psi_J(0,\hbar))\,+\,\E^{-itH(\hbar)/\hbar}\,\psi_J(0,\hbar)
\\[7pt]\nonumber
&=&\psi_{\tilde{J}(J,\hbar)}(t,\hbar)+
O(\|\E^{-itH(\hbar)/\hbar}
\psi_J(0,\hbar)-\psi_{\tilde{J}(J,\hbar)}(t,\hbar)\|)+
O(\|\psi_0(0,\hbar)-\psi_J(0,\hbar))\|).
\eea
Thus, to make the error terms exponentially small in $\hbar$
we need to consider values of the cutoff $J$ that grow to infinity
with $\hbar$ in a suitable way, and we also need to control our
approximation as a function of $J$.

In the proofs of all previous theorems, the dependence of the approximation
on $l$ governs the estimates on the error terms.
The $\hbar$ dependence comes through the different choices of $l\simeq g/\hbar$
or $l\simeq g(T)/\hbar$, with $T\simeq \ln(1/\hbar)$.
The set ${\cal C}$ is chosen to give an
exponentially small contribution as $l\ra \infty$ in the last
term of (\ref{strategy}) with the choice
\be\label{choiceJ}
  J=\nu\,l,
\ee
for some $\nu >0$.
We need only show that the basic estimates in the proofs
above are unaltered by the replacement of $J$ by $\nu\,g/\hbar$,
for $g$ small enough.

We can do this because we have been careful to make the $J$
dependence explicit in all the key estimates, such as
Corollary \ref{cor}.

In the contribution to the error term associated
with $\chi_1$ given by (\ref{explicitJ}) we adapt the last
step by using the estimate
\be\label{trick}
\frac{(J(l)+3l)!}{J(l)!}\ \leq\ c_0(\nu)\,\frac{((\nu+3)l)^{(\nu+3)l}}
{(\nu l)^{\nu l}}\ \leq\ c_0(\nu)\,c_1(\nu)^l\,l^{3l},
\ee
for some constants $c_0(\nu)$ and $c_1(\nu)$. Hence, the remainder of
the argument for Lemma \ref{inside} is the same, with updated constants. Since
the constants are modified in a time independent way, the long time estimates
are also unchanged.

Consider now the contribution associated with $\chi_2$ in Lemma \ref{outside}.
We first note that (\ref{choiceJ}) implies (with a slight abuse of notation)
$\tilde{J}(l)=J(l)+3l-3=(\nu+3)l-3$
so that we still have $\tilde{J}(l)\simeq g/\hbar$. The arguments
that rely on the smallness of $g$ to allow us to use
of Lemma (\ref{db-est}) remain in force. We thus arrive at
(\ref{explicitJ2}). We deal with it by using (\ref{trick}), exactly as above,
and obtain exponential decay again in case $l=g/\hbar$. The long
time estimates are also valid as the time dependence of the constants
is unaltered. This shows that Theorems \ref{thm1} and \ref{thm3}
are true with our generalized initial coherent states.

Theorem \ref{thm2} also holds for these initial states provided
we can control the sum in (\ref{explicitJ3}) with
$J(l)=\nu l$. To do so, we first note that the last two factors of
(\ref{explicitJ3}) can be bounded by $\E^{\beta' J}$, for some
$\beta'$, so that they are of order $\E^{g/\hbar}$.
This is harmless if $g$ is small enough because of the exponentially
decreasing prefactor. Next, we use $k\leq l-1$ to obtain
$$
\sqrt{(J+3k)!/(J!)}\ \leq\ (J+3k)^{3k/2}\,(J+3k)^{J/2}/\sqrt{J!}
\ \leq\ c(\nu)\,\frac{((\nu+3)l)^{\nu l/2}}{(\nu l)^{\nu l/2}}\,(J+3k)^{3k/2},
$$
for some constant $c(\nu)$.
Thus, the sum in (\ref{explicitJ3}) is bounded by
\be\label{bds}
c'(\nu)^{l}\ \sum_{k=0}^{l-1}\, C_2'^{k}\,\hbar^{k/2}\,(J+3k)^{3k/2}/k!,
\ee
for another constant $c'(\nu)$, and where $C_2'$ is proportional to
$C_0\,D_5$ in (\ref{explicitJ3}). Since
$$
(J+3k)^{3k/2}\ \leq\ {((\nu+3)l)^{3k/2}}\ \leq\
(((\nu +3)g)^{3/2}/\hbar^{3/2})^{k},
$$
we can bound (\ref{bds}) by
$$
c'(\nu)^{g/\hbar}\,\E^{C_2'((\nu +3)g)^{3/2}/\hbar}
$$
which, again, is harmless for sufficiently small $g$.

To prove the validity of Theorem \ref{thm4}, insert (\ref{cgt}) and
(\ref{ccgt}) in the estimates above and check that the conclusion
still holds. This is straightforward.

Finally, for Theorem \ref{thm5} to hold, we first must consider
the contribution associated to $\chi_1\,\xi_l$, which relies on  (\ref{first}).
Here, (\ref{trick}) applies directly. Next, the first contribution from
$\chi_2\xi_l$ is (\ref{Jexplicit4}). It has the form
(\ref{explicitJ3}) and yields exponential decay in the same way, for
$g$ small enough. It remains for us to bound (\ref{sumsum}).
With $s\,\leq\,l+1+2k\,\leq\, 3l$, we use the estimate,
$$
\sqrt{(J+s)!}\ \leq\ (\nu l+s)^{\nu l/2}\,(\nu l +s)^{s/2}\ \leq\
((\nu+3)l)^{\nu l/2}\,((\nu+3)g/\hbar)^{s/2}
$$
in (\ref{unesum}). The first factor when multiplied
by $\E^{\bar{\beta}J}/\sqrt{J!}$ is of order $\E^{cg/\hbar}$ where
$c$ is independent of $g$. The final factor allows us to repeat
the argument that led to (\ref{last}). Hence, for $g$ small enough, we
get an exponentially small contribution in $\hbar$ and the result
follows. \ep

\end{document}